\documentclass[iop,apj]{emulateapj}

\begin{document}

\newcommand{\kms}{\ensuremath{\mathrm{km}\,\mathrm{s}^{-1}}}
\newcommand{\galunits}{\ensuremath{\mathrm{km}\,\mathrm{s}^{-1}\,\mathrm{kpc}^{-1}}}
\newcommand{\galacc}{\ensuremath{\mathrm{km}^2\,\mathrm{s}^{-2}\,\mathrm{kpc}^{-1}}}
\newcommand{\MLsun}{\ensuremath{\mathrm{M}_{\sun}/\mathrm{L}_{\sun}}}
\newcommand{\Lsun}{\ensuremath{\mathrm{L}_{\sun}}}
\newcommand{\Msun}{\ensuremath{\mathrm{M}_{\sun}}}
\newcommand{\Aunits}{\ensuremath{\mathrm{M}_{\sun}\,\mathrm{km}^{-4}\,\mathrm{s}^{4}}}
\newcommand{\surfdens}{\ensuremath{\mathrm{M}_{\sun}\,\mathrm{pc}^{-2}}}
\newcommand{\voldens}{\ensuremath{\mathrm{M}_{\sun}\,\mathrm{pc}^{-3}}}
\newcommand{\gevcc}{\ensuremath{\mathrm{GeV}\,\mathrm{cm}^{-3}}}
\newcommand{\etal}{et al.}
\newcommand{\LCDM}{$\Lambda$CDM}
\newcommand{\ML}{\ensuremath{\Upsilon_*}}


\title{The Surface Density Profile of the Galactic Disk \\ from the Terminal Velocity Curve}

\author{Stacy S. McGaugh}
\affil{Department of Astronomy, Case Western Reserve University, Cleveland, OH 44106}
\email{stacy.mcgaugh@case.edu} 

\begin{abstract}
The mass distribution of the Galactic disk is constructed from the terminal velocity curve and the mass discrepancy-acceleration relation.
Mass models numerically quantifying the detailed surface density profiles are tabulated.
For $R_0 = 8$ kpc, the models have stellar mass $5 < M_* < 6 \times 10^{10}\;\Msun$, scale length $2.0 \le R_d \le 2.9$ kpc, 
LSR circular velocity $222 \le \Theta_0 \le 233\;\kms$, and solar circle stellar surface density $34 \le \Sigma_d(R_0) \le 61\;\surfdens$.
The present inter-arm location of the solar neighborhood may have a somewhat lower stellar surface density than average for the solar circle.
The Milky Way appears to be a normal spiral galaxy that obeys scaling relations like the Tully-Fisher relation, the size-mass relation,
and the disk maximality-surface brightness relation.
The stellar disk is maximal, and the spiral arms are massive.
The bumps and wiggles in the terminal velocity curve correspond to known spiral features (e.g., the Centaurus Arm is a $\sim 50\%$ overdensity).  
The rotation curve switches between positive and negative over scales of hundreds of parsecs. 
The rms amplitude $\langle$$|$$dV/dR$$|^2$$\rangle$$^{1/2} \approx 14\;\galunits$, 
implying that commonly neglected terms in the Jeans equations may be non-negligible.
The spherically averaged local dark matter density is $\rho_{0,DM} \approx 0.009\;\voldens$ ($0.34\;\gevcc$).
Adiabatic compression of the dark matter halo may help reconcile the Milky Way with the $c$-$V_{200}$ relation expected in \LCDM\ while also
helping to mitigate the too big to fail problem, but it remains difficult to reconcile the inner bulge/bar dominated region with a cuspy halo. 
We note that NGC 3521 is a near twin to the Milky Way, having a similar luminosity, scale length, and rotation curve.  
\end{abstract}



\section{Introduction}
\label{sec:Intro}

The structure of our Galaxy is notoriously difficult to discern given our location within it.  
The traditional picture of a disk plus bulge has progressed to include both thick and thin disks and a 
prominent bar component.  The thickened portion of the central bar
may account for much or perhaps even all of what was traditionally considered the bulge \citep[e.g.,][]{shen2010}.
Detailed models of the non-axisymmetric bar have been constructed \citep[e.g.,][]{Portail2015,MVO2015},
there has been enormous progress in mapping the vertical structure of the disk \citep[e.g.,][]{Binney2014,BienRAVE},
and the stellar halo is now known to contain considerable substructure \citep[e.g.,][]{helmi}.

Despite these advances, we persist in parameterizing
the radial surface brightness profile of the primary stellar component of the Galaxy as an exponential disk.
This is a crude approximation that ignores variations due to spiral structure, the kinematic effects of which have been 
detected \citep{ravespirals,Williams,Faure}.
Even with the simple exponential disk approximation, estimates of basic parameters like the scale length of the disk 
range from $R_d \approx 2$ kpc \citep[e.g.,][]{Gerhard2002} to 4 kpc \citep[e.g.,][]{glimpsescalelength}.

It would be good to move beyond the exponential disk approximation.  
Here we seek to supplement traditional photometric constraints on the stellar mass distribution 
with a different technique based on kinematic information.
The result is a numerical estimate of the stellar surface density profile $\Sigma_d(R)$.

A basic result from the mass modeling of external spiral galaxies is that features in the azimuthally averaged light profile 
have corresponding  ``bumps and wiggles'' in the rotation curve.  This can be phrased as Sancisi's Law: 
``For any feature in the luminosity profile there is a corresponding feature in the rotation curve and vice versa'' \citep{renzorule}.
This is quantified by the mass discrepancy-acceleration relation \citep[MDAR:][]{MDacc,GalRev}, which empirically
relates the baryonic mass distribution to the rotation curve.
We utilize this correspondence to infer features in the stellar surface density profile from those 
observed in the terminal velocity curve of the Milky Way. 

\section{Galactic Mass Models}
\label{sec:GMM}

The ideal map of the Galaxy would include complete 6D phase space information for every star.  
Within such a map one can imagine perceiving not just bulge and disk, or even thick disk and stellar halo, but a distinct
stellar population for each and every star forming event.  Chemical tagging \citep[e.g.,][]{ChemTag2009,ChemTag2010,Quillen2015} in the 
era of large surveys like GALAH \citep{GALAH} and Gaia \citep{GAIA} should help to move us closer to this ideal.

\begin{figure*}
\epsscale{1.0}
\plotone{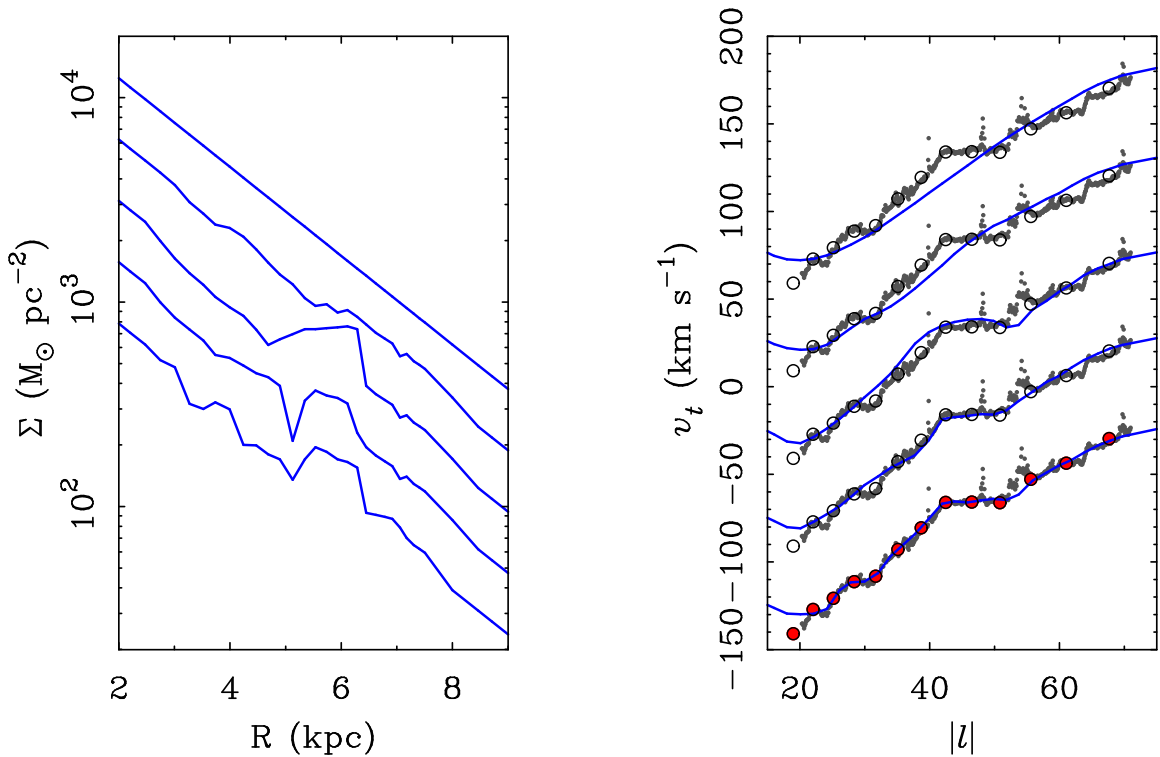}
\caption{Sequence of steps illustrating the terminal velocity fitting procedure.  
The disk surface density is shown in the left panel; the resulting velocities are shown in the right panel. 
We start with a smooth exponential disk model (straight line at top left) that provides what is usually considered an adequate 
description of the terminal velocity data [fourth quadrant HI data from \citet[gray dots]{MGD} and CO data from \citet[circles]{luna}].
We then redistribute mass in rings \citep{M08}, recompute the mass model, and compare it to the data.
Iterative steps are shown from top to bottom, ending when a satisfactory fit is achieved (bottom; CO data marked in red).
The correct y-axis values are shown for the final fit; previous iterations are offset by a factor of two in density (left) and 
by $50\;\kms$ in velocity (right) for clarity.
\label{fitseq}}
\end{figure*}

A desirable subset of the ideal map would be a 2D image $\Sigma(R,\phi)$ of the Milky Way 
as seen face-on by an external observer \citep{deVMW,GLIMPSE}.
At this juncture it is clear that such an observer would witness a strong bar in the central regions of the Milky Way.
Presumably spiral arms would be perceptible as well.  While there has been a great deal of recent work on the Galactic bar, 
the mass contained in the spiral arms remains uncertain.  Such features are certainly known to exist, both from star counts, 
and in the distribution of tracers in the $\ell$-$v$ diagram \citep{BM}.

Here we attempt to take one small step forward by applying what we have learned from external galaxies to the Milky Way.  
The product is a numerical, non-parametric representation of the azimuthally averaged surface density profile $\Sigma_d(R)$.
This is not simply a kinematic estimation of the disk scale length, as it includes the bumps and wiggles presumably induced by spiral arms.

\subsection{Assumed Galactic Parameters}
\label{sec:assume}

Our chief interest here is to find the relative variation in stellar surface density as a function of radius.
It is beyond the scope of this paper to address many of the outstanding problems in Galactic structure,
so we make some specific assumptions in order to move forward.  The absolute values of the surface densities
will likely need to be tweaked as more precise values of the Galactic constants are nailed down, but we expect 
the relative variations --- the bumps and wiggles of interest here --- will persist.

Specifically, we assume $R_0 = 8$ kpc and $\Theta_0 = 220\;\kms$.  
These set the scale of the rotation curve to which we fit.
We keep $R_0$ fixed but let $\Theta_0$ vary.  
The inferred variation is within the uncertainty in the solar motion.

Recent work indicates a slightly larger Milky Way \citep[e.g.,][]{Chatz2015}.
The relative variation of the bumps and wiggles are not strongly affected, and the absolute normalization of
the surface densities are only affected to a small degree.  
The total mass of the Galaxy varies with $R_0$, but not enough to alter any of the conclusions drawn here.
Perhaps the strongest effect of $R_0$ is on the shape of the rotation curve, which rises unnaturally if $R_0$ becomes too large.  
Consistency with the measured rotation curve shapes of external galaxies prefers $R_0 \lesssim 8$ kpc \citep{OM98}. 

\subsection{Method}
\label{sec:method}

The procedure is that described in \S 5 of \citet{M08}.
We start with a purely exponential stellar disk,
\begin{equation}
\Sigma_d(R) = \Sigma_d(R_0) e^{-(R-R_0)/R_d}.
\label{eq:expdisk}
\end{equation}
\citet{M08} found that the kinematic data preferred short disk scale lengths,
so for an initial guess we adopt $R_d = 2$ kpc
and a surface density at the solar ring $\Sigma_d(R_0) = 35\;\surfdens$ \citep{flynn}.
We compute the rotation curve of the stellar disk $V_d(R)$ using the GIPSY \citep{GIPSY} 
task ROTMOD, which numerically solves the Poisson equation for the stipulated mass distribution. 
An exponential vertical profile with scale height $h_z = 300$ pc \citep{Siegel2002} is assumed.

\begin{figure*}
\epsscale{1.0}
\plotone{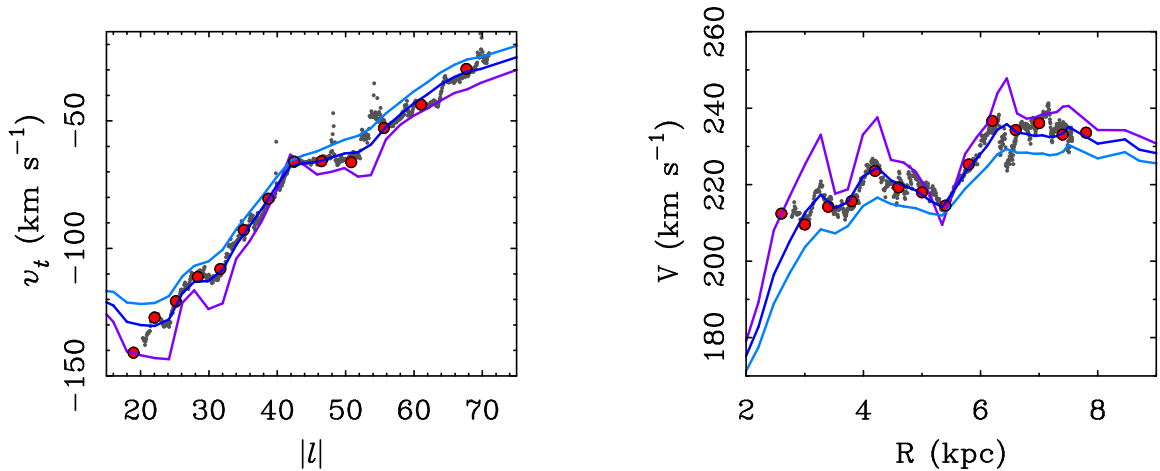}
\caption{The effect of disk thickness on the terminal velocities (left) and corresponding rotation curve (right).
Fits  (central line) have been made assuming an exponential vertical profile with $h_z = 300\;\mathrm{pc}$ \citep{Siegel2002}.  
The other lines show models with identical radial surface density profiles but different vertical scale heights.  A razor thin disk is shown
(line having more negative terminal velocities and higher rotation velocities) together with a thick disk with $h_z = 600\;\mathrm{pc}$ 
(line with less negative terminal velocities and lower rotation velocities).  
All other things being equal, thinner disks rotate faster and respond more strongly to bumps and wiggles in the surface density profile.
A disk of any plausible thickness could be made to fit the data by changing the mean surface density (which needs to be slightly higher for a 
thicker disk) and modulating the relative amplitude of the bumps and wiggles.  Though details may change for
different assumptions, the pattern of bumps and wiggles would remain the same.
\label{diskthickness}}
\end{figure*}

Next, we compute the corresponding baryonic rotation curve $V_b$ including the bulge ($V_B$) and gas ($V_g$):
\begin{equation}
V_b^2(R) = V_d^2(R) + V_B^2(R) + V_g^2(R).
\end{equation}
The treatment of the bulge is described in more detail below. 
The gas distribution is adopted from \citet{Olling}, including both atomic and molecular gas corrected for helium, as in \citet{M08}.

The baryonic rotation curve provides an estimate of the full rotation curve $V_c$ by way of the MDAR \citep[][]{MDacc,GalRev}:  
\begin{equation}
V_c^2 = D V_b^2.
\label{eq:Vtot}
\end{equation}
The MDAR is an empirical relation between the amplitude of the mass discrepancy $D$ and the force per unit mass $g_b = V_b^2/R$
generated by the baryons.  In effect, it quantifies Sancisi's Law.
To represent the MDAR we use
\begin{equation}
D = (1-e^{-\sqrt{g_b/a_{\dagger}}})^{-1}
\label{eq:MDAR}
\end{equation}
\citep[see][]{FBMW,M08,FM12}
with $a_{\dagger} = 3700\;\galacc$ \citep{BBS,MDacc,M11,M12,GalRev}.
This is the same functional form adopted by \citet{M08}.

The rotation curve computed in this way is compared to the observed rotation curve from the terminal velocities observed
interior to the solar circle.  The input stellar surface density of the disk $\Sigma_d(R)$ is adjusted by changing the surface density
in each ring by hand to match the features in the terminal velocity curve \citep[see \S 5 of][]{M08}.
This procedure is repeated as illustrated in Fig.~\ref{fitseq} until an adequate fit is obtained.

\subsection{Terminal Velocity Data}
\label{sec:tvdata}

We apply the method described above separately to the first and fourth quadrant terminal velocities.
For a given choice of ($R_0,\Theta_0$), the terminal velocities in the two quadrants are in effect separate realizations of the
Galactic rotation curve.  Differences between the quadrants may reflect 
real differences in the gravitational potential stemming from asymmetry in the mass distribution of the Galactic disk.
We take the terminal velocity at face value and derive corresponding $\Sigma_d(R)$ independently in each quadrant.  

First quadrant data are adopted from the CO observations of \citet{Clemens}, as these seem to underpin many published
estimates of the Galactic rotation curve.  Data for the fourth quadrant are taken from the CO observations of \citet{luna}
and the HI observations of \citet{MGD}.  These latter are the same data utilized in \citet{M08}.

\begin{deluxetable*}{llcccccccccccccc}
\tablewidth{0pt}
\tablecaption{Milky Way Models}
\tablehead{
\colhead{Model} & \colhead{B/T} & \colhead{$M_B$} & \colhead{$M_d$} & \colhead{$\Sigma_d(R_0)$} &
\colhead{$R_d$} &\colhead{$R_p$}& \colhead{$V_p$} & \colhead{$V_b$} & \colhead{$V_f$} &
\colhead{$\Theta_0$} & \colhead{$A$} & \colhead{$B$} & \colhead{$\ML^V$} & \colhead{$\ML^I$} & \colhead{$\ML^K$} \\
 & & \multicolumn{2}{c}{$10^9\;\Msun$} & \colhead{$\surfdens$} & \multicolumn{2}{c}{kpc} 
& \multicolumn{4}{c}{\kms} & \multicolumn{2}{c}{\kms kpc$^{-1}$} & \multicolumn{3}{c}{\MLsun}
}
\startdata
Q1ZB 	&0	&\phn0\phd\phn	&51.5	&35	&2.0	&6.1	&237	 &204	&204 &222	&13.8	&$-14.0$ 	& 1.38 & 1.20 & 0.60 \\
Q1MB 	&0.18    &10\phd\phn	&46.6	&35	&2.0	&6.1	&238	 &206	&205	 &224	&14.0	&$-14.0$ 	& 1.52 & 1.32 & 0.65 \\
Q1BB 	&0.30  &16\phd\phn        	&37.6	&34	&2.0	&6.3	&241	 &208	&208	 &224	&13.9	&$-14.1$ 	& 1.44 & 1.25 & 0.62 \\
Q4ZB 	&0	&\phn0\phd\phn	&55.1	&53	&2.4	&6.4	&239	 &205	&207 &232	&15.5	&$-13.4$ 	& 1.48 & 1.28 & 0.64 \\
Q4MB 	&0.21	&11.5        	&44.2	&53	&2.4	&6.4	&237	 &203	&207	 &232	&14.7	&$-14.3$ 	& 1.49 & 1.30 & 0.64 \\
Q4BB 	&0.34	&20\phd\phn	&38.2	&61	&2.9	&6.4	&236	 &201	&208	 &233	&13.9	&$-15.3$ 	& 1.56 & 1.36 & 0.67
\enddata
\tablecomments{The distance to the Galactic Center is assumed to be $R_0 = 8.0$ kpc.
The total gas mass in all models is $M_g = 11.8 \times 10^9\;\Msun$.
This includes both atomic and molecular gas, and has been corrected to include helium and metals.
The mass-to-light ratios assume that the total luminosity of the Milky Way is
$L_V = 37.3$, $L_I = 42.9$, and $L_K = 86.5 \times 10^{9}\;\Lsun$ \citep{drimmelspergel,flynn,Just2015}.}
\label{globaltab}
\end{deluxetable*}

The terminal velocity data provide an excellent tracer of the rotation interior to the solar radius through
\begin{equation}
V = v_t +\Theta_0 \sin \ell
\end{equation}
\citep{BM}, provided that the gaseous tracers are in circular motion at the tangent points.
This is a good approximation at larger radii, where the velocity dispersion is much less than the circular speed in both
stars and gas.  It breaks down as we approach the center of the galaxy and material becomes entrained in the Galactic
bar.  We fit the data over the range\footnote{The practical upper limit from which useful constraints come is $|\ell| \approx 70^{\circ}$
($R \approx 7.5$ kpc).} $3 < R < 8$ kpc ($22^{\circ} < |\ell| < 90^{\circ}$), with the understanding that the inner portion of 
this range may be affected by the bar. We do not fit the data within 3 kpc on the presumption that it certainly is.

Our procedure requires other judgement calls beyond the decision of where the eccentricities of orbits due to the bar
become too great.  We cannot hope to fit every tiny bump and wiggle.  Nor should we do so, as some may be due
to errors or non-gravitational (e.g., gas) physics.  Examination of the terminal velocity data reveal several qualitatively distinct features.
There are broad features extending over several degrees of Galactic latitude, the most prominent of which is that extending
from $\ell \approx -40^{\circ}$ to $-50^{\circ}$ in the fourth quadrant (Fig.~\ref{fitseq}).  These we fit.  There are also sudden, sharp deviations
in velocity that appear as sudden spikes in the HI data of \citet[e.g., at $\ell = -48^{\circ}$ and $-54^{\circ}$]{MGD}.  These we do not fit.
Indeed, no plausible mass distribution can explain such sudden changes in velocity.  We imagine that these features are
shocks or strong flows of gas where our line-of-sight crosses a spiral arm fragment.  Finally, there are intermediate features,
small and sometimes sharp but distinct from the sudden spikes.  These we fit as we can, without obsessing over 
differences of a few \kms\ that are smaller than turbulence in the gas \citep{kalberla09}.

The formal uncertainties in the terminal velocity measurements are typically only a few \kms.  
These are small compared to systematic uncertainties, particularly the degree to which the assumption of circular motion holds.
Since these are not quantified, we make no attempt at a formal fit that minimizes $\chi^2$.  Rather, we consider a fit to have
converged if the rotation curve passes through the bulk of the data, and captures the observed
pattern of bumps and wiggles.  Though we make no claim to have obtained a formal best fit, 
we are not aware of any other results that fit the terminal velocity data in such detail\footnote{The nearest comparable
work is that of \citet{sofuedip}, who fit a possible dip in the rotation curve outside the solar circle with a ring of mass further out.}.

\subsection{Disk Thickness}
\label{sec:diskthickness}

Considerable effort has been made to understand the vertical structure of the disk \citep[e.g.][]{Binney2014,Piffl2014,BienRAVE}.
Separate thin and thick disk components can be perceived, and these may have different radial scale lengths \citep{Juric2008}.
The exact vertical structure of the disk need not be purely exponential any more than its radial structure, and this is essential
to the determination of the vertical restoring force to the disk.

Here we are interested in the radial rather than the vertical force.  The vertical structure plays a relatively minor role in the
computation of $V_d^2/R$.  This is illustrated by Fig.~\ref{diskthickness}, which shows the terminal velocities and corresponding
rotation curve for three models of differing scale height.  The models are otherwise identical, sharing the same radial mass distribution
$\Sigma_d(R)$.  In addition to the nominal assumed scale height of 300 pc, a razor thin disk and a thicker disk with $h_z = 600$ pc
are shown. 

As expected \citep{BT}, the thinner disk rotates slightly faster and the thicker one more slowly, all other things being equal.
In addition, thinner disks respond more dramatically to variations in the surface density profile, also as expected.
However, the absolute difference between plausible models is not great.  We therefore fix the disk thickness 
to 300 pc \citep{Siegel2002} and do not distinguish between thick and thin disks.  For this particular problem, this distinction
is small, with differences that are smaller than those caused by turbulent motion in the gas.

\subsection{The Bulge-Bar}
\label{sec:bulgebar}

We are interested here in the structure of the Galactic disk, and in particular the detailed radial variation of its stellar surface density.
We make no attempt to infer this outside the solar radius, where the tangent point method cannot be applied, nor inside a radius of 3 kpc
($|\ell| < 22^{\circ}$) where non-circular motions are important. 
For the present purpose, the nature of the central component of the galaxy --- whether it is a bulge or a bar or some combination thereof --- 
is not terribly important.  However, it is necessary to account for the integrated interior mass.
To this end, we approximate the central ``bulge'' component with a numerical model based on the COBE light distribution \citep{BGS},
as described by \citet{M08}.  

Here we vary the normalization of the bulge component to check its effect on the inferred disk surface densities.  
As one might expect, the bulge plays only a minor role, and only at small radii.  
We build models with different bulge fractions to explicitly quantify its effect.

\section{Mass Models}
\label{sec:models}

We construct mass models with three components:  a stellar disk, a central bulge, and a gas disk.  
The gas disk is based on the work of \citet{Olling}, and is identical to that used in \citet{M08}.
The bulge model is also that used in \citet{M08}, but here we vary the bulge fraction, adopting three cases:
zero bulge, a nominal bulge fraction close to 20\% of the of the total light, and a heavy bulge that could be taken
to represent a bulge light fraction of $\sim 1/3$, or equivalently, a bulge with a smaller light fraction but
with a mass-to-light ratio heavier than that of the disk.  These cases presumably bracket reality.

A model is built for the stellar disk for each of the three choices of bulge fraction. 
This is done separately in the first and fourth quadrant, treating the terminal velocity curve from each as an independent estimate of the rotation curve.
This produces a total of six models.  The fourth quadrant model with zero bulge is basically identical to that
in Table 3 of \citet{M08}, with the exception that the subtle outward force of the gas component at small radii, ignored before, is treated rigorously here.

\begin{figure}
\epsscale{1.0}
\plotone{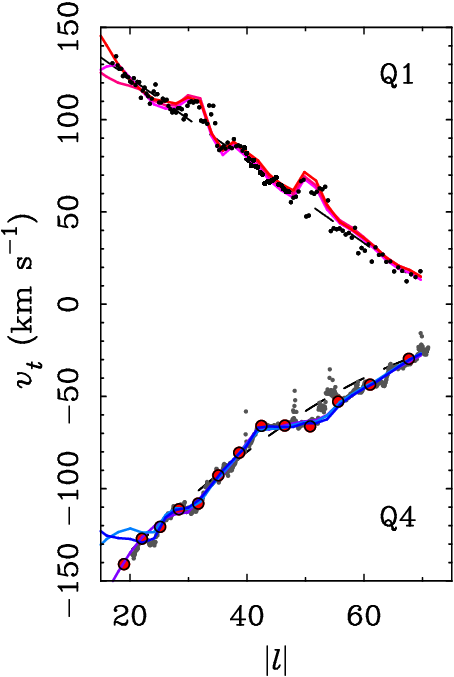}
\caption{Fits to the terminal velocity data for models with varying bulge fraction in the 
first (positive velocities) and fourth (negative velocities) quadrants.
The CO data of \citet{Clemens} are shown in the first quadrant.  In the fourth quadrant,
the CO data of \citet{luna} are shown as large circles and the HI data of \citet{MGD} as small dots.  
For reference, the smooth model of \citet{MGD} is shown as the dashed curve in the forth quadrant,
as is a simple linear fit to the first quadrant data ($v_t = 167.3-2.235 \ell$).   
The solid lines show our detailed fits to the data.  The models differ significantly only at small radii
($R < 3$ kpc; $|\ell| < 22^{\circ}$).  No attempt is made to fit
the data in this inner region where non-circular motions become important.  
Lines representing larger bulge fractions can be distinguished by their higher $|v_t|$.
\label{terminalV}}
\end{figure}

Fits to the terminal velocities are shown in Fig.~\ref{terminalV}.
The models are very similar over the range fit.
The effect of the different bulge fractions is apparent only at small radii ($| \ell | < 22$). 

The bulk properties of the models are given in Table~\ref{globaltab}.
All models assume a Galactocentric distance $R_0 = 8$ kpc.  
Scaling to other values of $R_0$ is not straightforward, but the basic pattern of bumps and wiggles would persist.

The first column of Table~\ref{globaltab} provides a label for each model composed of the 
the quadrant for which the terminal velocity data have been fit and the bulge size (Zero, Moderate, or Big).  
Q1 denotes the data of \citet{Clemens} while Q4 denotes those of \citet{luna} and \citet{MGD}.
The second column quantifies the bulge fraction of the total stellar mass.  
The third and fourth columns are the mass of the bulge and the disk in units of $10^9\;\Msun$.

The point of these models is to move beyond the usual approximation of an exponential disk.
Nevertheless, it is useful to fit an exponential disk to the inferred surface densities as a reference.
Equation~\ref{eq:expdisk} is fit over the range $3 < R < 8$ kpc where the terminal velocities have been fit.
The fifth and sixth columns give the surface density at the solar radius and the scale length that result from this fit.

It is interesting to see how the parameters of the fitted exponential disk vary.  
In the first quadrant, the bumps and wiggles average out, and return a fitted exponential
indistinguishable from the smooth initial guess.  In the fourth quadrant, a higher surface density is inferred at larger radii,
leading to fits with longer scale lengths and higher $\Sigma_d(R_0)$.  It is well known that such fits depend on the range over which
the fit is made.  That we obtain somewhat different results from the first and fourth quadrants may go some way to explaining the
range of results found in the literature, which may themselves be fit over different radial and azimuthal ranges.  

\begin{figure}
\epsscale{1.0}
\plotone{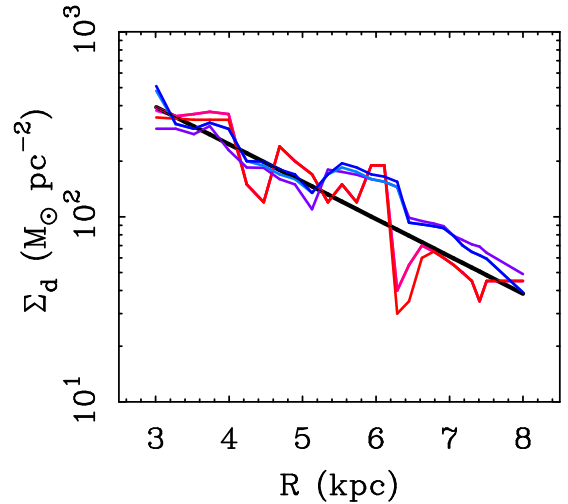}
\caption{The mass surface density of the stellar disk inferred from fitting the radial force as traced by the terminal velocities.  
Red lines are models fit to the first quadrant data and blue lines are those fit to the fourth quadrant data.
The straight black line is the surface density inferred independently from the vertical force by \citet{BR13}.
\label{MWsurfdens}}
\end{figure}

Another item to note is that the fitted surface density of the solar circle need not be identical to that of the solar neighborhood.
That is to say, our local patch extending over a small range of azimuths need not be identical to the ring centered at $R_0$ averaged over
all azimuths.  As it happens, the exponential disk fits to the bumps and wiggles in the first quadrant return a stellar surface density at the
solar radius very near to that measured locally \citep[e.g.,][]{flynn}.  In contrast, the fits to the fourth quadrant data indicate a rather higher
surface density.  This difference may just be fluke, but it may also indicate a real variation from one side of the Galaxy to the other.
Taken literally, it appears that the surface density averaged around all azimuths may exceed that of the solar neighborhood. 
That is, the sun might reside in a patch that is a bit under-dense for its radius, consistent with its current inter-arm location.

\begin{deluxetable*}{ccccccccc}
\tablewidth{0pt}
\tablecaption{Detailed Milky Way Mass Models}
\tablehead{
\colhead{Model} & \colhead{R} & \colhead{$\Sigma_d$} & \colhead{$V_d$} 
& \colhead{$\Sigma_{B}$} & \colhead{$V_{B}$} & \colhead{$\Sigma_{g}$} & \colhead{$V_{g}$}  & \colhead{$V_c$} \\
 & \colhead{kpc} & \colhead{\surfdens} & \colhead{\kms} & \colhead{\surfdens} & \colhead{\kms} 
& \colhead{\surfdens} & \colhead{\kms} & \colhead{\kms} 
}
\startdata
Q1ZB & 0.1      &2029    & 15.1      &       0    &    0     &      0   &   $-0.7$     &    20.5 \\
   & 0.2      &1930     & 29.3      &            &          &          &   $-1.5$     &    36.1 \\
    & 0.3      &1836     & 42.3      &            &          &          &   $-2.2$     &    49.9 \\
    & 0.4      &1746     & 54.4      &            &          &          &   $-3.0$     &    62.5 \\
    & 0.5      &1661     & 65.5      &            &          &          &   $-3.8$     &    74.0 
\enddata
\tablecomments{Table \ref{massmodelstable} is published in its entirety in the electronic edition of 
the Journal.  A portion is shown here for guidance regarding its form and content.
}
\label{massmodelstable}  
\end{deluxetable*}

Figure \ref{MWsurfdens} shows the surface density profiles of the models over the radial range to which exponential fits are made.
Models with different bulge fractions look similar, while those from different quadrants are notably different.
Also shown for reference is the stellar surface density profile found by \citet{BR13} [$\Sigma_d(R_0) = 38\;\surfdens$, $R_d = 2.15$ kpc]
from their analysis of the \textit{vertical} force.  This is in good agreement with that found here from the \textit{radial} force.

The seventh column of Table~\ref{globaltab} reports the radius at which the rotation curve of the baryonic mass model peaks.
This would be 2.2 scale lengths for a purely exponential disk, but can be different for general mass distributions.
For these models, the radius where $V_b$ peaks is a bit larger than $2.2 R_d$.
Column 8 is the total circular velocity at this radius, which is useful for making a fair comparison to external galaxies.
Column 9 is the rotation attributable to the baryons (stars and gas) at $R_p$.  All models work out to be essentially maximal
with $V_b/V_p \approx 0.85$.  This reflects the fact that within the solar radius, the Galaxy resides 
in a portion of the MDAR where the mass discrepancy is modest.

The tenth column gives the rotation velocity that an external observer would measure at large radii.
This is taken to be the velocity at the edge of the HI distribution at 20 kpc, where the HI surface density drops to the typical sensitivity
limit of $1\;\surfdens$.  This outer velocity is useful for comparing the Milky Way to other galaxies in the Tully-Fisher relation.  Note that
this $V_f$ is not equal to the circular velocity of the LSR, which is reported for each model in column 11.
All models have slowly declining rotation curves at large radii.  One consequence of this is that we cannot measure $\Theta_0$,
assume the rotation curve is flat, and expect this to be an adequate measure for comparison to external galaxies.

Columns 12 and 13 of Table~\ref{globaltab} give the Oort constants $A$ and $B$.  
These are determined from the local gradient in the rotation curve just inside and outside of the solar radius.
The models are not particularly well constrained at the solar radius, as the tangent point method is only effective interior to the solar circle.
The radial run of the Oort constants is discussed in \S \ref{sec:dvdroort}.

\begin{figure*}
\epsscale{1.0}
\plotone{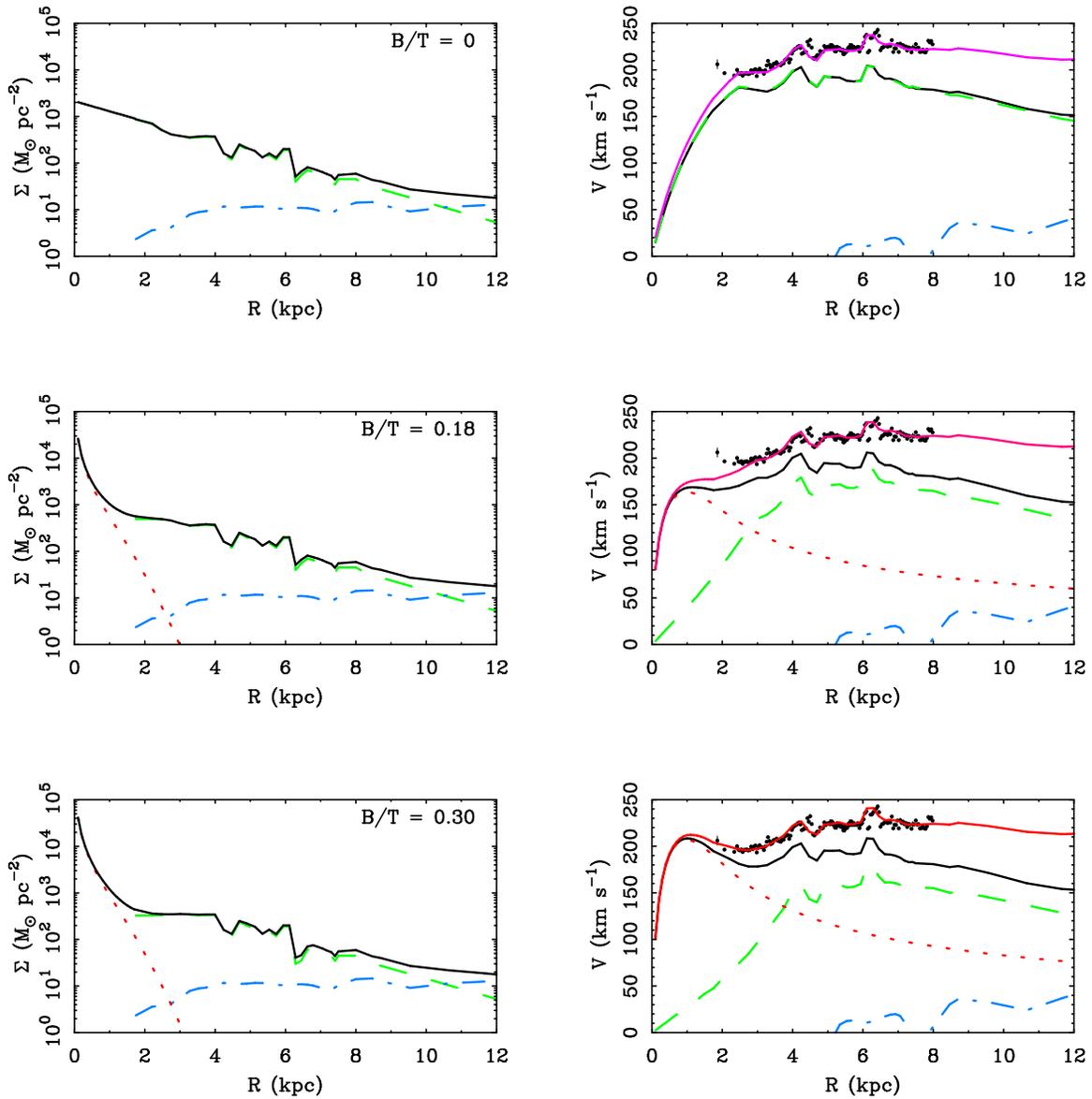}
\caption{Mass profiles (left) and rotation curves (right) for first quadrant models designated by bulge fraction.
Symbols as per Fig.~\ref{terminalV}.  
The stellar disk is represented by dashed lines, the bulge-bar by dotted lines,
and the gas disk by dash-dotted lines.  Their sum is given by the solid line.  The bumps and wiggles in the
mass profiles at left can cause the corresponding features in the rotation curve (upper solid lines in right panels),
as fit to the terminal velocity data of \citet[points]{Clemens}.
\label{massmodels1Q}}
\end{figure*}

\begin{figure*}
\epsscale{1.0}
\plotone{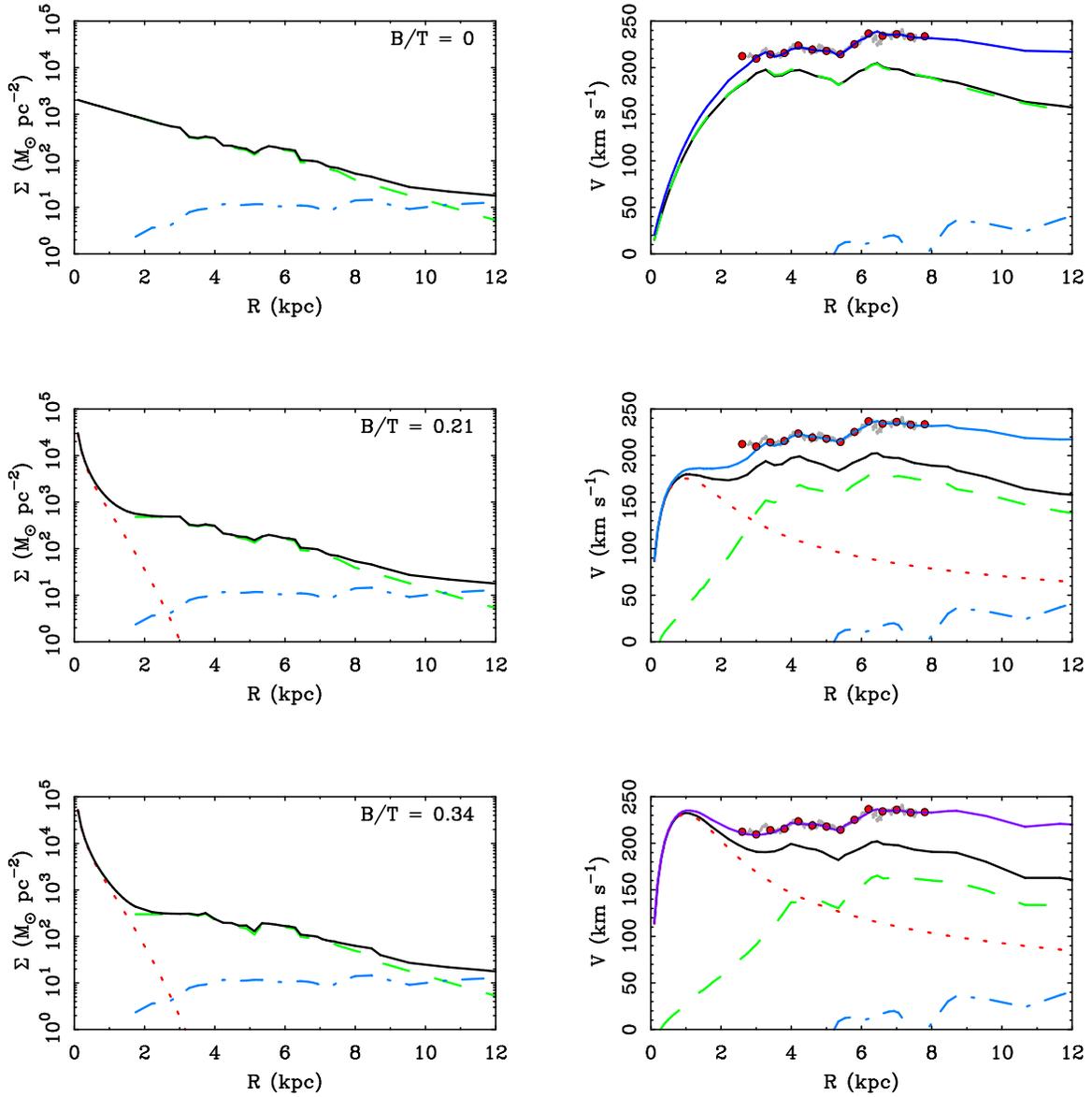}
\caption{Identical to Fig.~\ref{massmodels1Q} but for the fourth quadrant, where the models have been fit to the
data of \citet[red circles]{luna}.  The HI data of \citet{MGD} are also shown as small gray dots.
\label{massmodels4Q}}
\end{figure*}

The final three columns of Table~\ref{globaltab} give the stellar mass-to-light ratio of each model Milky Way in the $V$, $I$, and $K$ bands.
Integration of the models fit to the terminal velocities provides the stellar mass.  For the total luminosity of the Milky Way we adopt 
the $K$-band \textit{disk} luminosity of $6.92 \times 10^{10}\;\Lsun$ of \citet{drimmelspergel}, and follow their lead in correcting
this upwards to include a 20\% bulge fraction, resulting in a total luminosity of $L_K = 8.65 \times 10^{10}\;\Lsun$.
The other luminosities assume $V-I = 0.90$ and $V-K = 2.46$ which are \textit{local} colors from \citet{flynn} and \citet{Just2015}.
These yield $L_V  = 3.73 \times 10^{10}\;\Lsun$ and $L_I = 4.29 \times 10^{10}\;\Lsun$.  

Note that the total $K$-band luminosity
adopted here, after correction for the bulge, is very similar to the disk-only luminosity found by extrapolation of the local surface
brightness by \citet{Just2015} before inclusion of the bulge.  They adopt a longer scale length than found here, which may account
for part of this discrepancy.  However, it is not obvious that the results of \citet{drimmelspergel} and \citet{Just2015} can entirely be reconciled.
Indeed, \citet{Just2015} note that the single star Arcturus makes a substantial contribution to the local surface brightness measurement, 
so we remain cautious about how accurately these quantities are known.

The $K$-band mass-to-light ratio of all models is very nearly $\ML^K = 0.6\;\MLsun$, 
consistent with the expectations of population synthesis models \citep{MS14}.
Indeed, this value was adopted in the calibration of the MDAR \citep{GalRev}.
However, obtaining the same mass-to-light ratio for the Milky Way is not guaranteed, as the luminosity estimate is independent of
the stellar mass estimate.  That the models return a value consistent with the MDAR calibrated by external
galaxies provides some hope that the luminosity estimates are not too far off.  The mass-to-light ratios in $V$ and $I$ follow from the adopted colors,
and are also reasonable from the perspective of stellar populations \citep[compare to Table 7 of][]{MS14}.
\citet{LNB2015} quote a very similar $I$-band mass-to-light ratio to what we find here.
They find a redder global color than assumed here, leading to a correspondingly
higher $V$-band mass-to-light ratio.

In our models, the disk mass declines as the bulge fraction increases.  
This trade-off is necessary to keep the total mass in the right ballpark.
Indeed, to accommodate an increasing bulge fraction, it is necessary to reduce the mass of the disk in the inner regions.  
This is usually accomplished by letting the scale length of the disk grow \citep[e.g.,][]{flynn}.
However, it is no longer possible to fit the bumps and wiggles if we stretch out the disk too much.
To address this issue, we limit the disk mass reduction to the region of the bulge/bar
by adopting a \citet{freeman} Type II profile\footnote{More generally, allowing a break in the exponential disk
profile may help alleviate the tension between the mass of the bulge-bar and the scale length of the outer disk.} 
with a constant surface density region interior to a radius that depends
on the bulge fraction.  The zero models increase exponentially all the way to the center.  For $B/T > 0$, $\Sigma_d = \Sigma_c =$ constant
for $R < R_c$.  These values are tabulated in the detailed mass models given in Table~\ref{massmodelstable}.

Type II profiles are a common morphology for the surface brightness profiles of barred spiral galaxies.
One might imagine that the constant density region of the disk represents the azimuthally averaged bar, while the bulge is a thicker component.
However, no attempt has been made to match the details of either kinematics or photometry in the inner region where this trade-off is made.
Indeed, the choice of inner profile is rather arbitrary.  The parameters $\Sigma_c$ and $R_c$ are highly degenerate, 
and similar results could be obtained with different combinations.  Given this, and the uncertainty in the bulge itself, we simply
choose an $R_c$-$\Sigma_c$ pair that accommodates the target bulge fraction.  To fit the remainder of the rotation curve,
we fit the pattern of bumps and wiggles, allowing the exact bulge mass to vary slightly in order to also fit the amplitude of the rotation curve.
For this reason, the final bulge fraction need not be exactly the target fraction.  

Table~\ref{massmodelstable} gives the complete mass model for each case.
The first column specifies the model by quadrant and bulge fraction, as in Table~\ref{globaltab}.
The second column is the radius in kpc.
The third and fourth columns are the surface density of the stellar disk and the circular speed of the gravitational potential it generates.
Similarly, the fifth and sixth columns are the surface density and rotation curve of the bulge component, and the seventh and eighth columns
those of the gas disk.  Column 9 gives the total rotation curve of the model determined from equation~\ref{eq:Vtot}.
The model is extrapolated with an exponential disk well beyond the limits of the data.

The mass models tabulated in Table~\ref{massmodelstable} are shown in detail in Fig.~\ref{massmodels1Q} (first quadrant)
and Fig.~\ref{massmodels4Q} (fourth quadrant).  Bumps in the stellar surface density cause corresponding wiggles in the rotation
curve.  The pattern of bumps and wiggles is similar regardless of bulge fraction, but differs between the two quadrants.

Close comparison of the rotation curves in the two quadrants reveals that $V(R)$ is a bit higher at middle radii in the first quadrant
relative to the fourth quadrant, with the reverse being true at the edges of the data.  Blinking between the two resembles the flapping
of a bird's wings.  Since the quadrant midpoints are offset by $90^{\circ}$, this is consistent with the effects of an $m=2$ mode perturbation,
i.e., spiral arms.  

\section{Discussion}
\label{sec:discuss}

The results discussed above amplify the initial work of \citet{M08}. 
The pattern of features in the terminal velocity curve can be fit by a corresponding pattern of features in the surface density profile of the 
stellar disk.  Here we discuss some of the implications of these models.

\subsection{Spiral Structure}
\label{sec:spirals}

We have applied the MDAR to infer the azimuthally averaged stellar surface density implied by the pattern of bumps and wiggles
observed in the terminal velocity curve.  If these kinematically inferred features are real, then they
should correspond to physical structures.  These are presumably spiral arms.

\begin{figure*}
\epsscale{1.0}
\plotone{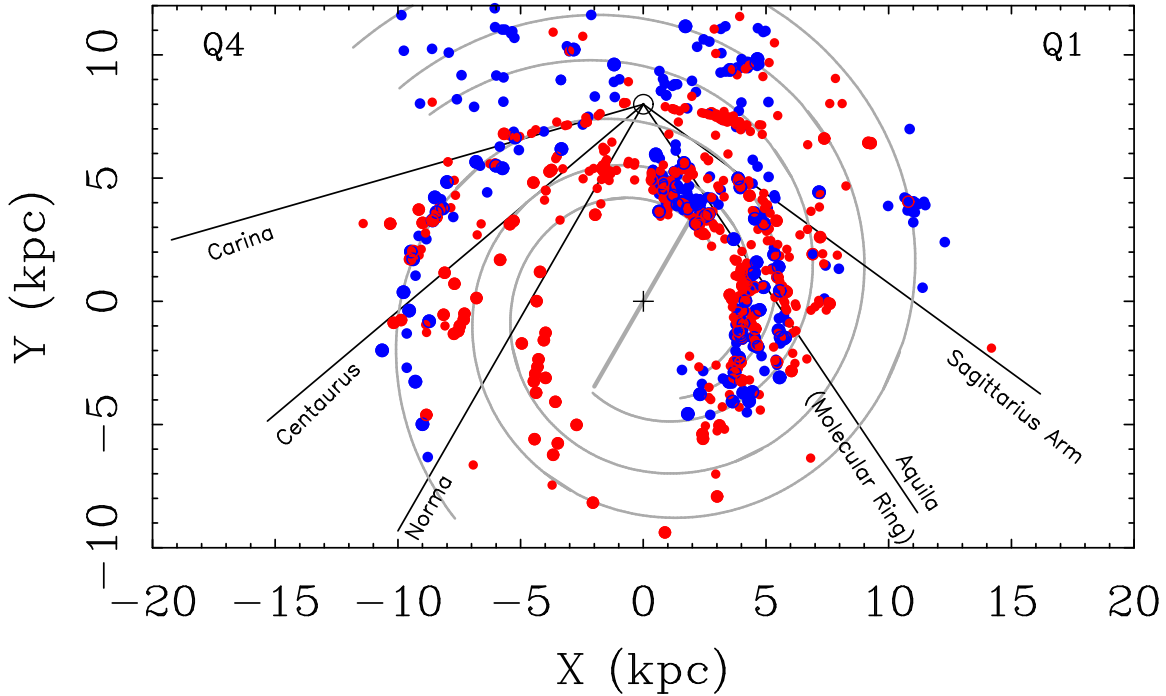}
\caption{The spiral structure in the Milky Way as traced by Giant Molecular Clouds (blue circles) and HII regions (red circles)
from the compilation of \citet{HHS}.  Small blue circles show GMCs with M$_{{GMC}} > 10^5\;\Msun$; large blue circles show 
those with M$_{{GMC}} > 10^6\;\Msun$.  Small red circles show HII regions with excitation parameters 
$U > 10\;\mathrm{pc}\,\mathrm{cm}^{-2}$; large red circles show those with $U > 100\;\mathrm{pc}\,\mathrm{cm}^{-2}$.  
For reference, the four arm spiral model of \citet{HHS} is shown (lines), as is a bar with a half length of 
4 kpc at a line-of-sight angle of $30^{\circ}$ \citep{weggbarangle}.  Lines of sight to the 
Carina, Centaurus, Norma, and Sagittarius arms are noted. 
Also noted is the line of sight toward Aquila which passes through the clump of material near the end of the bar and through the molecular ring.
Compare to Fig.~6 of \citet{deVMW}.
\label{spiralstruct}}
\end{figure*}

There is a rich history to the study of spiral structure in the Milky Way \citep{BM}.
Spiral arms are known features.  We can thus check whether the structures we infer correspond to known spiral arms.

Figure \ref{spiralstruct} shows the positions of known Giant Molecular Clouds and HII regions from the compilation of \citet{HHS}.
These objects are good tracers of spiral arms.  We also show lines of sight to known spiral arms:  the Carina, Centaurus, and Norma arms
in the fourth quadrant, and in the first quadrant the Sagittarius arm and the molecular ring in Aquilla.  This latter feature may simply be a tightly wound
spiral arm \citep{Dobbs2012}.

There is a good correspondence between known spiral arms and features in the terminal velocity curve.  
The most prominent is the Centaurus arm.  This manifests as the wide dip in the fourth quadrant rotation curve around $R \approx 5$ kpc
(Fig.~\ref{massmodels4Q}).  This can be seen in the surface density profile as a broad ($\sim 1$ kpc wide) overdensity extending a 
little beyond 6 kpc.

The Centaurus arm must represent a prominent overdensity of stellar mass to have the observed effect.  Bear in mind that the mass models
are axially symmetric, but fit to a single quadrant's data.  So, on the one hand, a bump must be large to have any effect on the azimuthally
averaged surface density profile.  On the other hand, the terminal velocities only probe the vicinity around the tangent point at each radius,
so features there may have an exaggerated effect.  Nevertheless, it is striking that spiral structure does appear to have the expected 
effect on the observed rotation curve.

Taking the data at face value, the Centaurus arm represents a $40\%$ enhancement over the smooth exponential fit of the corresponding models
in Table~\ref{globaltab}.  These fits include the overdensity.  Simply taking the ``background'' disk density as the line between the edges of the
Centaurus spiral feature raises the inferred enhancement to $60\%$.  This is an overdensity of mass associated with the Centaurus arm, 
not just light.

In the first quadrant, both the Sagittarius arm and the molecular ring/Aquila arm leave distinctive features in the terminal
velocities and corresponding surface densities.  Neither are as broad or massive as the Centaurus arm, but both have large density contrasts that
cause abrupt changes in the rotation curve.  These bumps and wiggles appear to be real features due to the expected effects of spiral structure.

The correspondence between features in the mass distribution inferred from the terminal velocity curves and known spiral arms
gives some confidence that the method employed here is on the right track. 
Just as the Milky Way has a central bar, so too it has spiral arms.  
These have the expected effect on the velocity field and the surface density profile of the stellar disk.
An obvious next step would be the construction of non-axisymmetric models $\Sigma_d(R,\phi)$.
These entail further degeneracies that are beyond the scope of this work.

\subsection{The Derivative of the Rotation Curve and the Oort Parameters}
\label{sec:dvdroort}

There are clear bumps and wiggles in the terminal velocity data.
These appear to correspond to real variations in the surface density of stars caused by spiral arms.
An interesting consequence is that there are non-negligible variations in the derivative of the rotation curve over scales of hundreds of parsecs.

\begin{figure*}
\epsscale{1.0}
\plotone{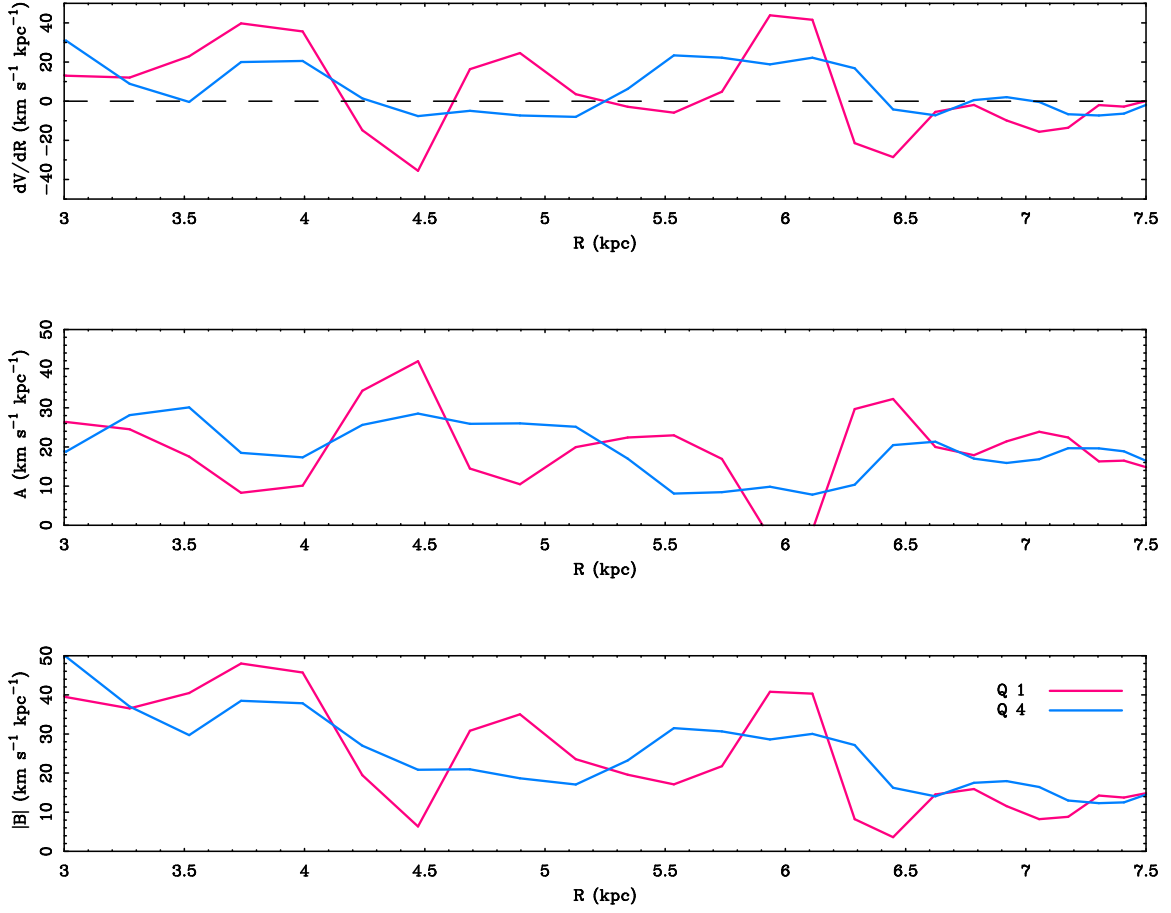}
\caption{The variation with radius of the derivative of the rotation curve $dV/dR$ (top panel), the Oort $A$ parameter (middle panel),
and the absolute value of the Oort B parameter (bottom panel).  
These quantities are illustrated for models Q1MB (red lines) and Q4MB (blue lines).  
Other bulge fractions are similar.  While the rotation curve is
approximately flat on average ($dV/dR = 0$: dashed line in top panel), there are large and apparently real excursions, both positive 
and negative, at all radii that are constrained by the data.  Consequently, the values of the Oort parameters that are determined by a
given survey will depend on how that survey samples these bumps and wiggles.
\label{dVdR}}
\end{figure*}

To a first approximation, the rotation curve is roughly flat over many kpc. 
Extrapolation of the models to larger radii predict a slowly declining rotation curve. 
However, on scales of hundreds of parsecs, $V(R)$ can switch between rising and falling rather suddenly.  
These local changes in $dV/dR$ can have important effects on the determination of quantities that depend on it (Fig.~\ref{dVdR}).

Examples of quantities that depend on the derivative of the rotation curve include the Oort constants and the vertical restoring force of the disk.  
The definition of the Oort constants depends explicitly on $dV/dR$, so their inferred value will differ between surveys that cover different
radii and azimuths if these happen to encounter different bumps and wiggles.  This difference will be manifest even if everything else
is done perfectly, so apparent conflicts between different data sets may instead represent real variations in the Galaxy.  
Rather than being a smooth function of radius, $A(R)$ and $B(R)$ can have quite a bit of structure (Fig.~\ref{dVdR}).

Fig.~\ref{dVdR} shows the radial variation of the derivative of the rotation curve and the Oort constants implied by the models.
The plot is restricted to the radial range over which the derivative can be extracted from the models fit to the terminal velocities,
which does not extend to the solar radius.  The models smooth out at $R>7.5$ kpc, but this is simply due to the end of the ability
of the data to constrain $dV/dR$.  In order to constrain the derivative at the solar radius, we need 
information beyond it that the terminal velocities do not provide.  Presumably the bumps and wiggles do not end where we happen to be.  
Indeed, the presence of the Perseus arm slightly outside the solar radius \citep{ravespirals} 
presumably provides another bump that wiggles the rotation curve beyond the solar radius \citep{sofuedip}.

\begin{figure*}
\epsscale{1.0}
\plotone{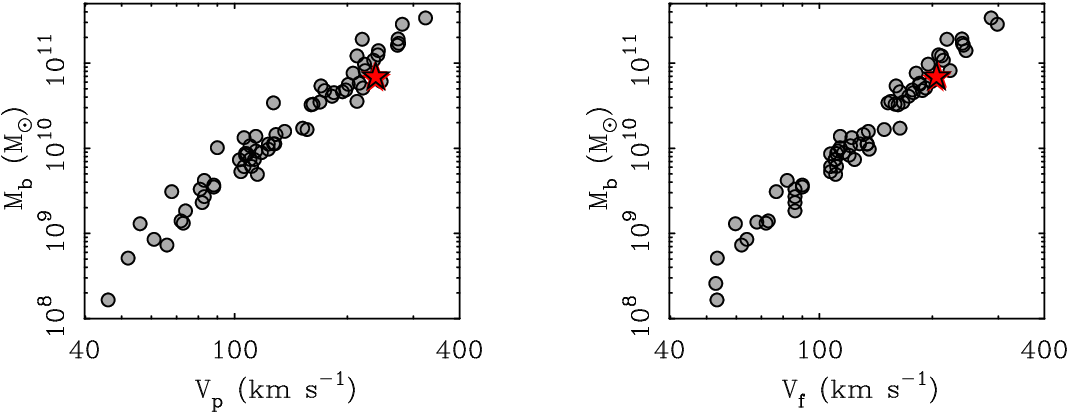}
\caption{The Milky Way on the baryonic Tully-Fisher relation, as illustrated by
galaxies from \citet{MS15} and from \citet[][excluding duplicates]{M05}.
Rotation velocities $V_p$ in the left panel are measured at the peak of the baryonic rotation 
curve \citep[left panel, equivalent to $V_{2.2}$ for a pure exponential disk]{myPRL} and the flat outer
velocity $V_f$ in the right panel.  For the Milky Way, $V_f$ is estimated as the model velocity at the last point
of the HI disk at $R=20$ kpc.
All six Milky Way models are plotted as red stars, with a single model outlined in black.  
The models are very nearly identical in this plane, differing by less than the symbol size.
The Milky Way resides within the small scatter in this relation: it appears to be a normal spiral galaxy in this context.
\label{BTFMW}}
\end{figure*}

The derivative of the rotation curve swings between positive and negative several times over the range depicted in Fig.~\ref{dVdR}.
The frequency with which this happens depends somewhat on how we choose to fit the observed bumps and wiggles in the terminal velocities.  
There is more variation than we have chosen to fit, though as discussed in \S \ref{sec:tvdata}, the smaller scale fluctuations are less
likely to by dynamical structures.  In the current models, the sign of $dV/dR$ changes over scales of hundreds of parsecs, with
the actual zero crossing being even more sudden.

To quantify the amount of variation in the derivative, we compute its rms over the range depicted in Fig.~\ref{dVdR}.
In the first quadrant model, $\langle$$|$$dV/dR$$|^2$$\rangle$$^{1/2} = 22\;\galunits$.
In the fourth quadrant, $\langle$$|$$dV/dR$$|^2$$\rangle$$^{1/2} = 14\;\galunits$.
For comparison, $V/R \approx 30\;\galunits$ at $R=7.5$ kpc, so the variation in the derivative is not much smaller than the orbital frequency
just interior to the solar neighborhood.

If the fine-grained rotation curve is not smooth, analyses that simplify the Jeans equations by assuming $dV/dR = 0$ may be incorrect,
or at least run the risk of introducing systematic errors.  For example, a non-zero rotation curve gradient on hundreds of parsec scales
might help to explain mild inconsistencies in the vertical force estimated over several kpc \citep{BR13}.
If this effect is important, one might expect it to manifest as apparent discrepancies within different subsets of the Gaia data.
That is, as these data become available, analyses that assume $dV/dR = 0$ may give different results when applied to 
subsets of the data representing distinct regions with different local gradients.
Imposing a uniform assumption about the gradient may lead to perplexing results.

\subsection{The Milky Way in the context of External Spirals}
\label{sec:context}

An obvious question is how the Milky Way compares to other spiral galaxies.
From the perspective of the Copernican Principle, one would expect it to be a normal spiral galaxy.
Occasionally, one finds cause to think it peculiar, which must also be true at some level of detail
since all objects are individuals.  Here we use the models constructed above to place the Milky Way in context.

Figure \ref{BTFMW} shows the Baryonic Tully-Fisher relation with the Milky Way highlighted.
All six Milky Way models from Table~\ref{globaltab} are plotted together with data for other galaxies from \citet{M05} and \citet{MS15}.
The sum of the baryonic mass components is plotted against two measures of the rotation velocity: that measured at the peak of the
baryonic contribution to the rotation curve [$V_p = V(R_p)$ being the generalized version of $V_{2.2}$], and that in the
outer, more nearly flat portion of the rotation curve ($V_f$).

The Milky Way, as modeled here, falls within the scatter of the Tully-Fisher relation for either measure of the rotation speed.
Indeed, all six models lie comfortably within the scatter, and are hardly distinguishable in the the Tully-Fisher plane.
There is perhaps a hint that the Milky Way lies on the lower right side of the very small scatter, but this is well within
the uncertainties.  Similarly, adopting a different value of $R_0$ will vary the rotation velocity and mass, but not beyond
the scatter for plausible values of $R_0$.  By this standard, the Milky Way is a normal spiral.

\begin{figure*}
\epsscale{1.0}
\plotone{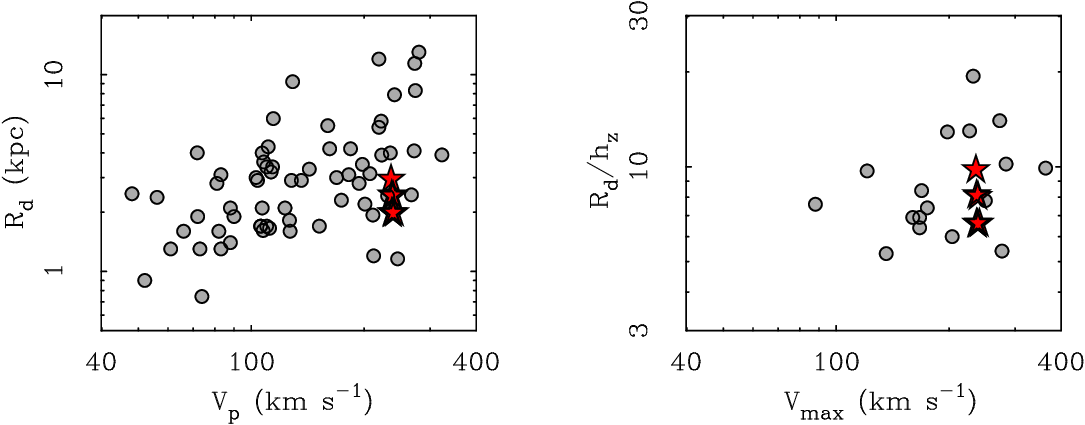}
\caption{The scale lengths (left) and thicknesses (right) of disks as a function of rotation speed.
The data in the left panel are the same as those in Fig.~\ref{BTFMW}.
The data in the right panel are from \citet{kregel1} and \citet{kregel2} and show the ratio of exponential
disk scale length to scale height.  
Differences between the models is more apparent here than in Fig.~\ref{BTFMW}.
The Milky Way does not stand out
from either distribution, though it is somewhat compact for its rotation velocity.
\label{Rdhz}}
\end{figure*}

Figure \ref{Rdhz} shows the disk size-mass and disk thickness-mass relation for spiral galaxies.
In both cases, the rotation speed is used as a proxy for mass.  This is a good proxy (see Fig.~\ref{BTFMW}),
but slightly different measures of the velocity are available: $V_p$ in the left panel, and the maximum observed velocity
$V_{max}$ in the right panel.  There is a strong correlation between $V_p$ and $V_{max}$, so these suffice here.
The exponential scale length is used to characterize the sizes of disk galaxies in the left panel of
Fig.~\ref{Rdhz}, which uses the same data as in Fig.~\ref{BTFMW}.  
For disk thickness in the right panel, data for edge-on galaxies are taken from \citet{kregel1} and \citet{kregel2}.

As with the Baryonic Tully-Fisher relation, the Milky Way appears to be a normal spiral galaxy.
It is perhaps a bit compact for its rotation speed, but this is within the ample scatter in the size-mass relation.
Segregation between the models of Table~\ref{globaltab} is more apparent here.  This is also true in terms of disk thickness,
where $h_z = 300$ pc is assumed for models of all radial scale lengths.  Nonetheless, the Milky Way sits comfortably
within the scatter shown by spiral galaxies generally.

The global properties inferred for the Milky Way by the modeling of the terminal velocities made here suggests that
the we live in a fairly ordinary spiral galaxy.

\subsection{NGC 3521: a Milky Way Twin}
\label{sec:ngc3521}

In comparing the Milky Way to other galaxies, we noticed that it not only follows the same correlations as other spirals,
but that it consistently falls in the same spot as one other galaxy in the comparison sample.  This object is NGC 3521.

NGC 3521 has a baryonic mass, rotation velocity, and scale length that are very similar to those of the Milky Way.
Indeed, these quantities are identical within the uncertainties.  While it is common for galaxies to
fall in the same spot along the Tully-Fisher relation, such Tully-Fisher pairs of galaxies often have
different disk scale lengths \citep{dBM96,TVbimodal,CR,myPRL}.  The intrinsic scatter in the
Tully-Fisher relation is small while that in the size-mass relation is large (see Figures \ref{BTFMW} and \ref{Rdhz}).

\begin{figure*}
\epsscale{1.0}
\plotone{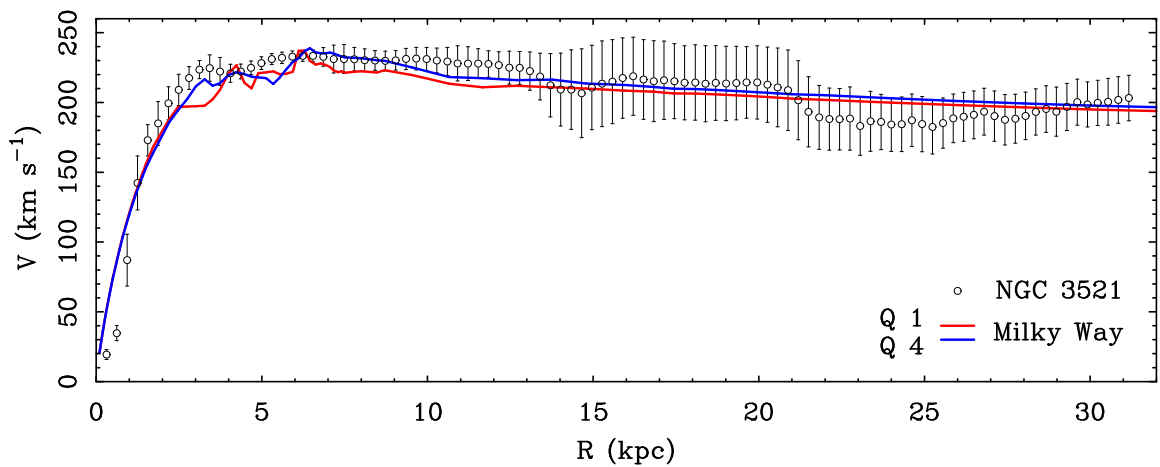}
\caption{The rotation curve of NGC 3521 \citep{THINGS} compared to the B/T = 0 Milky Way models Q1ZB (red line) and Q4ZB (blue line).
The two galaxies are near twins, having similar masses, disk scale lengths, and rotation curves.
\label{twins}}
\end{figure*}

In the case of NGC 3521, the scale length is close to that of the Milky Way as well as its position on the Tully-Fisher relation.
Indeed, the similarity persists in even greater detail.  Figure \ref{twins} shows the rotation curve of NGC 3521 from the THINGS
survey \citep{FTHINGS,THINGS}.  Also plotted are the zero bulge models of the Milky Way from both the first and fourth
quadrants (Table \ref{massmodelstable}).  The rotation curves of the two galaxies are practically indistinguishable.

There is a general lesson here.  
Galaxies that share a similar baryonic mass distribution share a similar rotation curve \citep{URC,GalRev}.
Nature builds galaxies from a very strict recipe \citep{Disneysimple}.  

Galaxies with structural similarities to the Milky Way have been noted before:
\citet{deVMW} highlighted NGC 1073, NGC 4303, NGC 5921, and NGC 6744.
These galaxies have similar global properties and morphological classifications.
Apparently NGC 3521 can be added to this list.
\citet{deVMW} classified the Milky Way as SAB(rs)bc. NGC 3521 is classified as SABbc \citep{RC3}.
Apparently both galaxies contain bars \citep[see also][]{NGC3521bar}, though the similarity of the inner regions is less pronounced.

\subsection{Maximal Disks}
\label{sec:maxdisk}

The disk inferred here for the Milky Way is maximal.  The baryons dominate the gravitational potential interior to the sun 
and contribute the bulk of the rotational support at $R_p$.  

\begin{figure*}
\epsscale{1.0}
\plotone{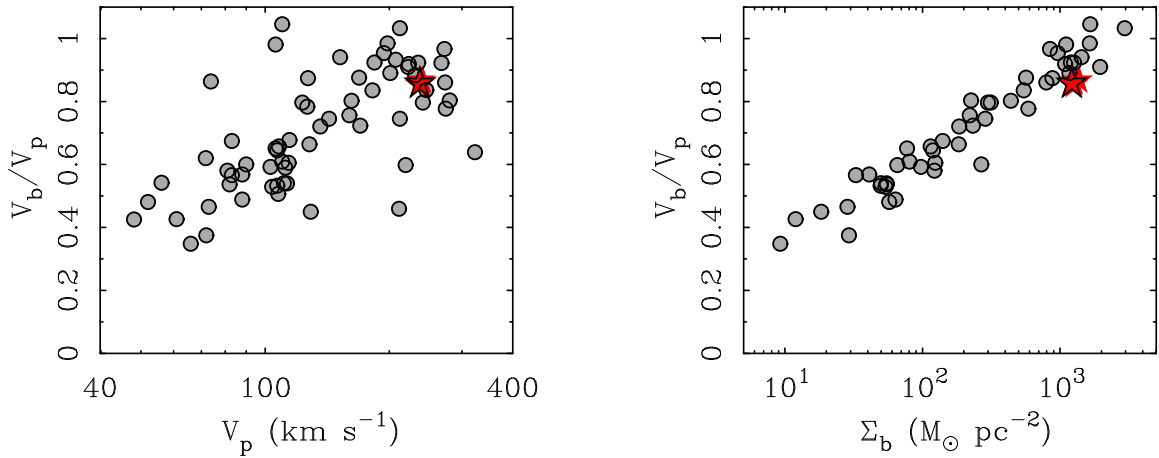}
\caption{The maximality of disks as a function of rotation speed (left) and characteristic baryonic surface 
density \citep[defined as $\Sigma_b = (3M_b)/(4R_p^2)$ by][right]{myPRL}.
The comparison data are the same as those in Fig.~\ref{BTFMW}.
The Milky Way has a maximal disk, as expected for a high surface brightness spiral \citep[see also Fig.~23 of][]{BR13}.
\label{MWmaxVbVp}}
\end{figure*}

One working definition of a maximal disk is one in which the
ratio of the rotation due to the baryonic component to the total rotation at $R_p$ is $V_b/V_p \approx 85\%$ \citep{sackett}.
The Milky Way models constructed here all fall in the range $0.85 < V_b/V_p < 0.87$.  
It therefore appears that the Milky Way has a maximal disk.  
This is consistent with the findings of \citet{sackett} and
\citet{BR13}, and with the expectation for a galaxy of the inferred surface density \citep[Fig.~\ref{MWmaxVbVp}]{myPRL}.

Indeed, it is necessary for the disk to dominate the gravitational potential in order to generate the bumps and wiggles in the terminal velocity curve.
Such features are the natural consequence of disk self-gravity \citep[e.g.,][]{SellwoodRev}. 
They cannot be supported by the dynamically hot, quasi-spherical dark matter halo \citep{BT}.  
That the procedure applied here works suggests that the the disk is massive.

The importance of disk self-gravity means that spiral arms are overdensities in mass, not just light.
If spiral arms were not enhancements in mass, then there would be no reason to expect an association between them
and features in the terminal velocity curve.  Since the expected imprint of massive spiral arms on the velocity field is observed,
it follows that the they are indeed massive.

It also follows that the disk in which spiral arms are embedded is also massive.
If we seek to reduce the disk mass, we are obliged to increase the contrast of the overdensity that the arms represent.
This is not small to start, and becomes greater as the surface density of the disk declines.
In effect, the spiral arms must become more massive as the disk becomes less massive.
This places an effective lower limit on the maximality of the disk that depends only on how unreasonable a contrast we are willing to tolerate.
The estimate of the contrast here is already relatively large \citep[cf.][]{drimmelspergel}.

\subsection{The Implied Dark Matter Distribution}
\label{sec:DM}

The implied distribution of dark matter follows trivially once the baryon distribution is specified.
The portion of the velocity attributable to the dark matter halo can be found from the data in Table~\ref{massmodelstable} through
\begin{equation}
V_{DM}^2 = V_c^2 - V_b^2.
\label{eq:DM}
\end{equation}
More generally, this can be expressed analytically as
\begin{equation}
V_{DM}^2 = V_b^2 (D-1)
\label{eq:MDACC}
\end{equation}
\citep{MDacc,GalRev}.
The amplitude of the mass discrepancy increases outwards as the centripetal acceleration decreases.
In the solar neighborhood the mass discrepancy $[\Theta_0/V_b(R_0)]^2$ is $D_0 \approx 1.5$.
At $R_p$,  $D_p = [V_p/V_b(R_p)]^2 \approx 1.35$.  At smaller radii, $D \rightarrow 1$ so that it becomes
difficult to perceive the role of dark matter in the inner few kpc \citep{Portail2015}. 

Once the functional form of the MDAR is specified, the rotation curve due to the dark matter halo follows.
This is only as uncertain as the empirical calibration of the MDAR and the mass-to-light ratio of a given galaxy.  
The density of dark matter can be inferred from $V_{DM}$ through solution of the Poisson equation.
Assuming a spherical halo,
\begin{equation}
4 \pi G \rho_{DM} = 2 \left(\frac{V_{DM}}{R}\right) \left(\frac{\partial V_{DM}}{\partial R}\right) + \left(\frac{V_{DM}}{R}\right)^2.
\label{eq:DMdensity}
\end{equation}
Applying this to the solar neighborhood, we infer a spherically averaged dark matter density of
\begin{equation}
\rho_{0,DM} \approx 0.009\;\voldens = 0.34\;\gevcc.
\label{eq:DMdens}
\end{equation}
This quantity is of obvious interest to laboratory searches for dark matter, and is very similar to other 
estimates \citep[e.g.,][]{HF2000,SNGFM2010,McMillan2011,BT12,strigariMW,Read2014JRev,PifflRAVE,Piffl2014,Piffl2015}.
Indeed, it is so consistent with previous results that it did not warrant mention in \citet{M08}, though the same result can be derived from
the information provided there.  It is equally trivial to derive the dark matter density at any other point in the Galaxy by combining
equations \ref{eq:MDACC} and \ref{eq:DMdensity}.

The dark matter profile dictated by equation \ref{eq:MDACC} does not, in general, follow any of the traditional analytic prescriptions 
for dark matter halos.  We can nevertheless fit such halo models, which give a tolerable description of the data over the modest
range of radii probed.  For example, a pseudo-isothermal halo characterized as
\begin{equation}
V_{ISO}(R) = V_{\infty} \left[1-\left(\frac{R_C}{R}\right) \arctan\left(\frac{R}{R_C}\right)\right]^{1/2}
\end{equation}
fits the data with a core radius $R_C = 3$ kpc and an asymptotic velocity $V_{\infty} = 177\;\kms$ (Table~\ref{halotable}).

Similarly, the NFW halo \citep{NFW} is characterized by a concentration $c$ and a characteristic velocity $V_{200}$.
This is the orbital velocity of a test particle on a circular orbit at the quasi-virial radius $R_{200}$.
This quasi-virial radius contains a mass density 200 times the critical density of the universe, and is typically far beyond 
the reach of observation. We can nevertheless use this notional quantity to define a radial variable $x = R/R_{200}$,
and the concentration $c = R_{200}/R_s$, where $R_s$ is the scale radius where the density profile rolls over \citep{NFW}.
The rotation curve of an NFW halo is 
\begin{equation}
V_{NFW}(R) = V_{200} \left[\frac{\ln(1+cx)-cx/(1+cx)}{x[\ln(1+c)-c/(1+c)]}\right]^{1/2}.
\end{equation}
Fitting this to the dark matter distribution indicated by the MDAR over the radial range $3 < R < 8$ kpc 
gives $c = 5.2$ and $V_{200} = 264\;\kms$ (Table~\ref{halotable}).

The NFW halo fit directly to the dark matter distribution given by the MDAR  
implies a rather large mass for the Milky Way of $M_{200} \approx 6 \times 10^{12}\;\Msun$.  However, our fit
over the radial range $3 < R < 8$ kpc has little power to constrain the total mass of the halo.  Indeed, NFW halos are highly self-degenerate,
so that a correlated series of $c$-$V_{200}$ values yield nearly indistinguishable results over finite ranges of 
radii \citep[hence the banana shaped contours in Fig.~4 of][]{dBMR2001}.  Consequently, an adequate description of the data is
also provided by an NFW halo with $c = 7.5$ and $V_{200} = 180\;\kms$.  Though not the formal best fit, this is well within the uncertainties. 
The total mass of this halo is $M_{200} = 1.9 \times 10^{12}\;\Msun$,
more in line with observational determinations from tracers at large radii \citep{MWhalomass}.

As a consequence of the degeneracy between $c$ and $V_{200}$, it is possible to fit still lower mass halos ($\lesssim 10^{12}\;\Msun$).
A frequently noted advantage \citep[e.g.,][]{skinnyMW} of a low mass Milky Way
halo is that it eases the so-called too big to fail problem \citep{2B2F}.  
This comes at the price of excessively high concentrations ($c \approx 20$).  These are not consistent with the predictions
of \LCDM, which provides a well defined mass-concentration relation \citep{MDvdB2008}.  
Simply lowering the mass of the Milky Way halo does not provide a satisfactory solution as the amount of substructure
depends on density, not just mass \citep{ZB2003,MBdB03}:  a high concentration is just as bad as a high mass in this context.

The mass-concentration relation is predicted by dark matter-only simulations \citep{NFW}.
One inevitable effect of forming a luminous galaxy within a dark matter halo is adiabatic compression.
This has the effect of raising the effective concentration of the resulting halo above that of the primordial initial condition to which the 
mass-concentration relation applies.

\begin{figure}
\epsscale{1.0}
\plotone{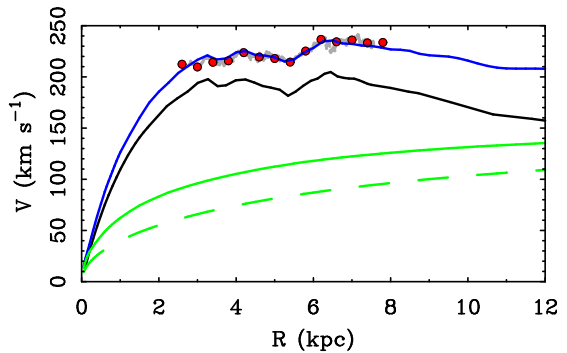}
\caption{The Milky Way rotation curve from the data of \citet{luna} and \citet{MGD} as in Fig.~\ref{massmodels4Q}
together with model 4QZB (black line). 
The total rotation is approximately fit (blue line) with an adiabatically compressed NFW halo (solid green line)
using the procedure implemented by \citet{adiabat}.  The pre-compressed, primordial halo is shown as the dashed line.
\label{DMdist}}
\end{figure}

We use the procedure of \citet{adiabat} to make an approximate fit to model 4QZB.
Both the primordial and compressed halo are illustrated in Fig.~\ref{DMdist}.
The compressed halo is the one subject to observational constraint, and no longer has a purely NFW form.
We can nevertheless fit it as if it were, with the resulting parameters given in Table~\ref{halotable}.
The result is a rather low mass halo ($M_{200} = 4 \times 10^{11}\;\Msun$) with a
concentration ($c = 14$) that is too high for \LCDM\ \citep{MDvdB2008}.
However, this is not the right result to compare to the prediction of simulations, which provide the mass-concentration relation prior to compression.
The pre-compressed, primordial halo is found to have a mass $M_{200} = 6.1 \times 10^{11}\;\Msun$ with a concentration $c = 7.1$.
This is nicely consistent with the predicted mass-concentration relation.

This result appears quite favorable.  Not only is the Milky Way consistent with the mass-concentration relation once we have accounted for 
adiabatic compression, the mass is low enough with a reasonable concentration to help with the too big to fail problem.
However, it would be premature to call this a complete solution, as the problem extends beyond the virial radius of the Milky Way \citep{GKBKBK14}.

There is another problem with the model in Fig.~\ref{DMdist}.
Close examination reveals that the rotation velocity is slightly over-fit at small radii, and under-fit at large radii.
This is a consequence of the cuspy center of the NFW halo, which predicts more mass at small radii and less at large radii than indicated
by the data \citep{M07}.  While the fit illustrated in Fig.~\ref{DMdist} is within the uncertainties, this problem rapidly exacerbates as we
allow for a non-zero bulge component.  As the mass distribution of the stars becomes more concentrated than the inward extrapolation of
a pure exponential disk, the compression of the halo becomes more severe.  The inner rotation curve becomes unrealistic at fairly modest bulge fractions.
This problem comes as no surprise, as it has been seen before \citep{Dubinski94,Abadi03}.  The compression of an initially cuspy dark matter halo
by a dense stellar bulge predicts much more dark matter at small radii than tolerated by observations \citep{Portail2015}.  We only escape this
problem in Fig.~\ref{DMdist} because we have made no attempt to model the inner 3 kpc.  \citet{Piffl2015} similarly conclude that compression in the
inner regions needs to be counteracted in some way.

\begin{deluxetable}{lcc}
\tablewidth{0pt}
\tablecaption{Dark Matter Halos}
\tablehead{
\colhead{Halo Model} & \colhead{$c$} & \colhead{$V_{200}$ (\kms)} 
}
\startdata
NFW MDAR & \phn5.2 & 264 \\
Compressed NFW & 14.0 & 107 \\
Primordial NFW & \phn7.1 & 124 \\
& $R_C$ (kpc) & $V_{\infty}$ (\kms) \\
Pseudo-isothermal & \phn3\phd\phn & 177
\enddata
\label{halotable}  
\end{deluxetable}

Note that it does not matter if the inner mass concentration is a bulge or a bar: the compression occurs in either case \citep{adiabat}.
While bulge formation may be chaotic, the adiabatic assumption remains fairly good \citep{Choiadiabat}.
The secular growth of a bar within the disk is the poster child for an adiabatic process.  Dynamical friction of the bar against the 
halo \citep{Debattista2000,athanassoula2003} may transfer angular momentum to the halo, 
but this is unlikely to counteract the compression \citep{Selwood2008}.
We can, as always, invoke feedback, but it is far from obvious that this will have the desired effect.
We therefore urge caution in the interpretation of Fig.~\ref{DMdist} and Table~\ref{halotable}, which appears promising but leaves important
questions unaddressed.



\subsection{The Predicted Vertical Force}
\label{sec:Kz}

The surface density of baryons can be constrained by the vertical restoring force to the disk \citep{KG89b,BR13}.
This provides an independent check on the surface densities derived here from the radial force.
More generally, the dark matter density inferred from the vertical force may differ from the spherical average given in \S \ref{sec:DM}
if the halo is oblate or if there is a distinct ``dark disk'' in addition to the quasi-spherical halo \citep{Read2014JRev,SSSRB15}.
It is therefore interesting to compare the results derived from vertical and radial force analyses.

The vertical force is given by the baryonic surface density and a term containing $\partial V^2/\partial R$ that emerges from the Poisson equation
as in equation \ref{eq:DMdensity}.  This ``tilt'' term can be cast in terms of the Oort parameters so that the vertical force $K_z$ is given by 
\begin{equation}
K_z = 2 \pi G (\Sigma_d + \Sigma_B + \Sigma_g) + 2Z(A^2-B^2).
\label{eq:vertforce}
\end{equation}
We can use the information from Table~\ref{massmodelstable} to predict $K_z$.

\begin{figure*}
\epsscale{1.0}
\plotone{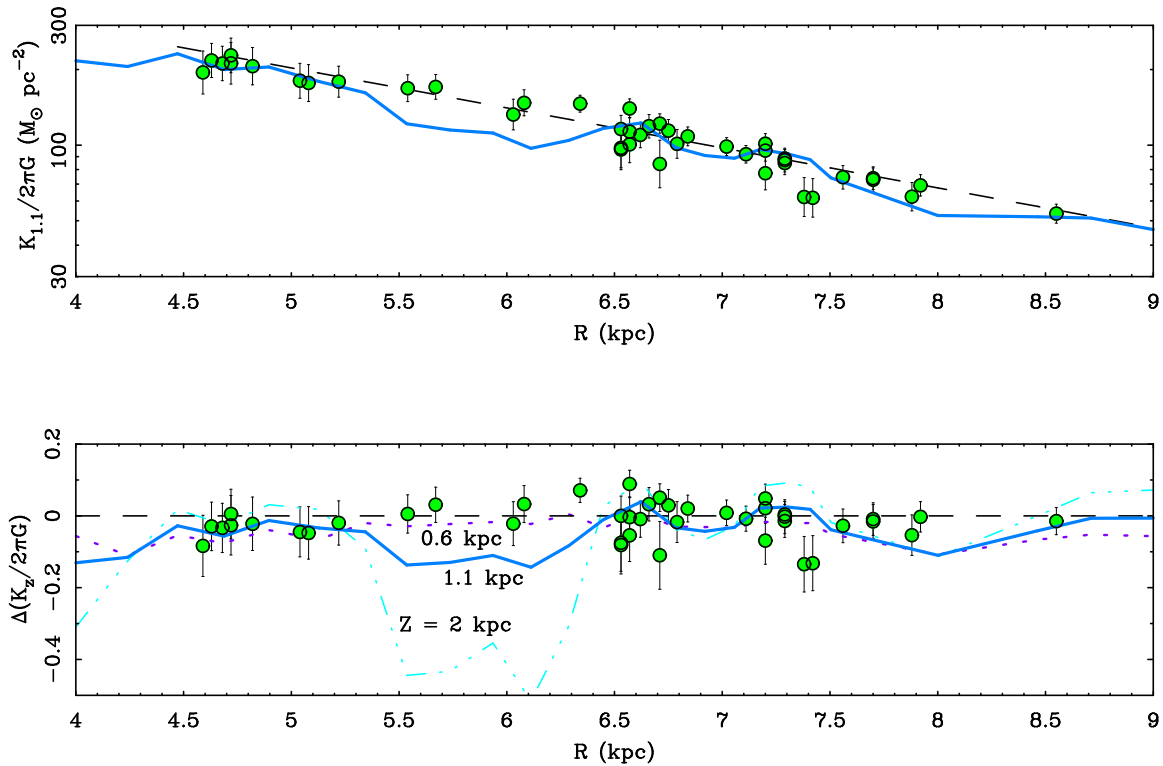}
\caption{Top panel: the vertical restoring force to the disk predicted by model 4QMB (solid line) compared to the
data of \citet[points]{BR13}. The data were measured at $Z = 1.1$ kpc, which is where the model is evaluated.  
The exponential fit of \citet{BR13} is shown by the dashed line, which is also depicted in Fig.~\ref{MWsurfdens}. 
Bottom panel: the vertical force relative to the exponential fit of \citet{BR13} (dashed line).
To illustrate the sensitivity of the vertical force to the tilt term $2Z(A^2-B^2)$, we also show lines for $Z = 600$ pc (dotted) and 2 kpc (dash-dotted).
\label{MWKz}}
\end{figure*}

The predicted vertical force of model 4QMB is shown in Fig.~\ref{MWKz}.
Also shown are the data of \citet{BR13}.
Other fourth quadrant models give similar results.  First quadrant models have more variation, but follow the same trend.

From examination of Fig.~\ref{MWKz}, it is clear that the models constructed here from the radial force and the MDAR are
compatible with the vertical force measured by \citet{BR13}. Indeed, the fourth quadrant models provide very nearly as good a fit
to these data as does the exponential fit made by \citet{BR13}.  This occurs despite the fact that the two quantities are completely independent.
The force predicted by our models follows from equation \ref{eq:vertforce} with no adjustment.
Indeed, this is a true prediction, as the essential information already appears in Table 3 of \citet{M08}.

Looking in detail at Fig~\ref{MWKz}, the only place where the model does not provide a good description of the data is over the one kpc range
$5.5 < R < 6.5$ kpc.  This is where the contribution of the $(A^2-B^2)$ is greatest (Fig.~\ref{dVdR}) due to the Centaurus spiral arm.
Elsewhere, the match is impeccable.

The difference between model and data may be real, as the two probe different regions of the Galaxy.
The data used by \citet{BR13} are located primarily along $\ell \approx 0$ (Bovy 2013, private communication).
The fit to the terminal velocities follows the locus of tangent points.  This locus deviates most from $\ell = 0$ precisely
at the mid-latitudes where the model and data disagree.  If the feature in the terminal velocities is indeed due to the Centaurus spiral arm,
one would expect it to shift in radius with azimuth, and perhaps change in amplitude as well.  Thus the one apparent failing of the model
may actually contain further information about Galactic structure.

The vertical force depicted in the top panel of Fig.~\ref{MWKz} is evaluated at $Z = 1.1$ kpc.
Equation \ref{eq:vertforce} depends linearly in $Z$ on the $(A^2-B^2)$ term.  This tilt term can become very important
when the derivative of the rotation curve is large.  To illustrate this sensitivity, in the bottom panel we illustrate the relative force
evaluated at $Z = 600$ pc and $Z = 2$ kpc.  While these retain the same basic shape as the $Z = 1.1$ kpc value at most radii,
their amplitude changes quite a bit where the tilt is large.  

Variation in the tilt has important implications for attempts to evaluate the vertical force
in order to determine the local dark matter density \citep[e.g.,][]{BienRAVE,silverwoodlocalDM}.  Normally, one assumes that
$\partial V^2/\partial R = 0$, or at least that it varies slowly and continuously.  This does not appear to be the case (Fig.~\ref{dVdR}).
Instead, the tilt term can vary substantially from place to place, and change suddenly.  This might go some way to reconciling
contrary analyses [e.g., \citet{monibidin} and \citet{BT12}], but it is dreadfully inconvenient.  If this effect is significant, it may manifest 
as apparent inconsistencies within the Gaia data when one assumes no fluctuations in the derivative of the rotation curve.
It also appears that the odds are good that many tracers could be affected: there are fluctuations 
everywhere \citep{wobblyMW,AntojaRAVE,XuRings}.

\subsection{The Relation to MOND and the Nature of Dark Matter}
\label{sec:MOND}

We have used the MDAR \citep{MDacc} to connect the terminal velocity curve to the surface density of the Galactic stellar disk.
With the calibration of the MDAR with external galaxies \citep{GalRev}, this is a purely empirical exercise.
This approach is valid irrespective of the physics underlying it.
What works for other spirals works for the Milky Way.

How the MDAR comes to be remains a mystery.
In the context of \LCDM, we are obliged to imagine that this very uniform scaling relation 
somehow emerges from the chaotic process of feedback during galaxy formation.
Alternatively, it could be that the appearance of a universal \textit{effective} force law in galaxy data (the MDAR)
is an indication of an actual modification of the force law \citep[MOND:][]{milgrom83}.

In the conventional dark matter context, the baryons are embedded in a dark matter halo.
There is no way to attempt the exercise successfully performed here because there is too much freedom in the dark halo model.
If we start by assuming an exponential disk, we never get to the point of fitting the bumps and wiggles, as we must first
fix the scale length.  This is effectively impossible, as there is degeneracy between the dark matter halo and baryon distribution:
one can shorten or lengthen the disk scale length to accommodate one halo model or another.  If, for example, we start by
assuming an NFW halo --- a very reasonable starting point in the context of \LCDM\ --- then we are immediately pushed towards
adopting longer scale lengths for the reasons discussed in \S \ref{sec:DM}.  As a consequence of the central cusp in NFW halos,
we never approach the maximum disk limit, and never suspect that the bumps and wiggles might be connected to the sub-dominant
baryon mass.  If instead we start with a maximum disk, then we are driven to overestimate the stellar surface density at large radii to
explain the bumps and wiggles near the solar circle.  This leaves little room for dark matter at small radii, so we may be inclined to adopt a halo model
with a low density density core.  This can fit the data, but we have no cosmological context for the halo parameters and remain afflicted with considerable
degeneracy between them and the baryons.  
In short, this exercise has not been done before because it is not possible without the MDAR.


The MDAR was uniquely anticipated by \citet{milgrom83}. 
The mapping between terminal velocities and features in the baryon distribution is very natural in MOND \citep{M08}.
It is not natural in the context of dark matter.

That said, the results for the vertical force may pose a problem for MOND.
The vertical force is computed conventionally in \S \ref{sec:Kz}.
The tilt term in equation \ref{eq:vertforce} is derived from the normal Poisson equation, not a modified version thereof \citep{BM84}.
The derivative of the rotation curve thus implicitly assumes that dark matter is the reason why the total rotation exceeds that 
predicted by Newton for the baryons.  As can be seen in Fig.~\ref{MWKz}, this works quite well.

In MOND, the Newtonian prediction for the radial force is amplified as the acceleration decreases below
the critical value, $a_0$ \citep{milgrom83}.  One would naively expect the vertical force to be enhanced by the same factor as the radial force.
The result is to make the dynamical scale length longer in MOND than it is conventionally \citep[by a factor of $\sim 1.25$:][]{Bienayme}.
Nevertheless, the conventional computation of the vertical force is consistent with a purely Newtonian maximal disk plus the dark matter halo
specified by the MDAR: it has not been stretched by the factor predicted by MOND (but see below).  

The match to the vertical force using the conventional formula is very good (Fig.~\ref{MWKz}).
It would seem like an extraordinary coincidence that this should occur by accident. 
By the same token, this can also be said for MOND fits to rotation curves in general:  
it is hard to imagine that this is a coincidence devoid of physical meaning.

It is worth noting that the Milky Way is not unique in this apparent mismatch between radial and vertical forces in MOND.
\citet{angusdiskmass} analyzed the vertical velocity dispersions of the face-on galaxies of the DiskMass survey \citep{diskmassI}
in the context of MOND.  In general it is possible to obtain a fit to the rotation curve or to the vertical velocity dispersion profile,
but not to both simultaneously.  In the case of the DiskMass survey, the thickness of each disk is not observed directly, so it is 
possible\footnote{A further complication is that the conventional analysis of the DiskMass data \citep{DiskMass7} obtains sub-maximal disks,
contrary to the results here.  The mean DiskMass mass-to-light ratio is a factor of $\sim 2$ lower than that anticipated by stellar population
models \citep{MS14,MS15}.  One possibility is that the stellar population that dominates the lines from which the vertical velocity dispersion 
is measured has a different scale height than the bulk of the stelar mass (e.g., the fractional contribution to the spectra by red supergiants 
can be larger than their contribution to the mass).  Another possibility is that we are not yet in a position to rigorously explain the vertical
force in disks, even conventionally.}
to obtain a simultaneous fit, but this only comes at the expense of making the disks much thinner than indicated by 
edge-on samples of similar morphological types.
Intriguingly, the shapes of the profiles for both the radial and vertical forces are well predicted, but the amplitude is offset.
It is as if MOND is more active in the radial direction than in the vertical direction: $D(Z|R) < D(R)$.

One interpretation is that the MDAR is simply an empirical scaling law, and does not embody new physics.
This is tempting, but leaves unanswered why it occurs in the first place. 
It is also tempting to dismiss MOND entirely for this discrepancy, and that might be the correct thing to do.
We should bear in mind, however, that we did not get this far without it: 
this apparent failing of MOND is not a success of \LCDM.
Indeed, it might be considered a success if a \LCDM\ galaxy formation model came within a factor of 1.25 of 
matching the scale lengths inferred independently from the radial and vertical forces, if such a test could ever be made.
Moreover, MOND does correctly predict \citep{Bienayme} the tilt angle of the velocity ellipsoid \citep{Siebert2008}.

MOND makes a number of definitive predictions \citep{roadtomond}.  
A precise mapping between radial and vertical force is not one of them.
For this, we need a specific version of the theory: 
one can construct MOND theories either by modifying gravity \citep[e.g.,][]{BM84,QUMOND} or 
inertia \citep{mondannals,mondasmodifiedinertia}.  
The increase of the scale length by a factor of 1.25 over the Newtonian expectation is specific to the formulation of \citet{BM84}.
In the case of modified inertia, the theory is inevitably non-local \citep{mondannals},
with the consequence that the dynamics is trajectory dependent.  So it is conceivable that the modest apparent mismatch between
radial and vertical forces in MOND is telling us something further about the correct underlying theory.

Another possibility to consider is something else entirely.  For example,
\citet{Blanchet} have proposed a type of dipolar dark matter that reproduces MOND's successes in galaxies while preserving those of \LCDM\ 
on larger scales.  Similarly, a dark matter superfluid \citep{Khoury2015} may possibly explain galaxy dynamics while behaving differently
in clusters of galaxies \citep{DMsuperfluid}. 
It is unclear at present what these theories predict for the vertical force of disks.

What is clear is that it is important to explore new ideas, with an open mind as to whether the mass discrepancy problem
is caused by some form of [not necessarily cold] dark matter or a modification of dynamical laws.
A minimum requirement for a successful theory is to explain the observed coupling between baryons and dynamics.
It is far from obvious that one can reasonably hope that the unique effective radial force law observed in galaxies
embodied by the MDAR will somehow
emerge from the chaotic feedback processes widely invoked in the context of cold dark matter.

\subsection{Hobgoblins of Inconsistency}
\label{sec:hobgob}

The models presented here are intended to provide a first step forward from the simplistic assumption of an exponential stellar disk.
As discussed above, they have a number of virtues, such as a realistic radial mass distribution that correlates with observed spiral structure
and is consistent with independent data.  However, they are not a complete solution, and we should be aware of a number
of minor respects in which they are not self-consistent.

The models constructed here are azimuthally symmetric. 
Though the radial mass distribution $\Sigma_d(R)$ is not smooth as in the usual exponential approximation,
we have made no attempt to construct two dimensional models $\Sigma_d(R,\phi)$.
Yet the bumps and wiggles that we identify in the surface density profile appear to correspond to spiral arms,
which certainly vary in azimuth.  Indeed, this is apparent in the difference between the first and fourth quadrant
terminal velocity curves, which imply bumps and wiggles at different radii.  This is expected as the stellar mass
follows the pitch angles of spiral arms to different radii as the azimuth varies.  
In effect, the models provide a snap shot of $\Sigma_d(R)$ along the azimuths probed by the terminal velocities.

In external galaxies, the full velocity field is observed, and a true azimuthal average is made.
This is what goes into the MDAR. 
In the Milky Way we only sample the rotation curve in the first and fourth quadrants, and not at all on the side opposite
the Galactic center.  Consequently, local variations in the surface density can have a larger impact on 
the rotation curves deduced from the terminal velocities than they might over a complete azimuthal average.
This probably results in the features inferred here being stronger than they would be after azimuthal averaging,
emphasizing the importance of local features.

The likelihood of azimuthal as well as radial variations further complicates the prospects for Jeans analyses.
In addition to the radial variation in $\langle$$|$$dV/dR$$|^2$$\rangle$$^{1/2}$ discussed in \S \ref{sec:dvdroort},
one should also worry about $dV/d\phi$ and how disk structure varies around the disk \citep{OD2003}.  
The Milky Way disk could be grand design or a patchwork of flocculent spiral structure,
rendering the usual azimuthally symmetric, radially smooth exponential disk approximation 
inadequate for the analysis of complex data like that provided by Gaia.

The terminal velocity curves in the first and fourth quadrants are different, and this difference appears to reflects a real
difference in the structure of the Galaxy.  This difference leads to differences in the first and fourth quadrant models.
These difference lead to rather different inferences for the local disk surface density $\Sigma_d(R_0)$ and LSR velocity $\Theta_0$.
The models fit to the first quadrant terminal velocities are quite consistent with the assumed (solar neighborhood) inputs for these values.
The fourth quadrant models prefer larger values of both $\Sigma_d(R_0)$ and $\Theta_0$.
This may indicate that the solar neighborhood is a bit underdense relative to the azimuthal average for $R_0$.
This is quite reasonable: there is no reason for the local patch around the sun to be exactly average, and the sun is known
to currently reside in a low density inter-arm region. 

A greater difficulty arises from the difference in $\Theta_0$.  
A particular value of $\Theta_0 = 220\;\kms$ has been assumed (\S \ref{sec:assume}) in order to derive the rotation curve.
The fits to the rotation curve then imply a slightly larger value for $\Theta_0$.
We should therefore iterate the solution, changing the rotation curve and re-fitting the surface densities.
In practice, this makes little difference to the result, and falls within the uncertainties of solar motion and gas turbulence,
and indeed of the variation in structure that we are attempting to ascertain.

First quadrant models give a circular velocity for the LSR of $\Theta_0 \approx 224\;\kms$, similar to the assumed value.  
For the observed proper motion of Sgr A* of $30.24\;\galunits$ \citep{ReidSGR}, this implies a rather high solar motion of
$V_{\sun} \approx 18\;\kms$.  While not inconceivable \citep{bovycrazy}, this is not consistent with 
the low solar motion found from the terminal velocity data themselves by \citet[$V_{\sun} = 7\;\kms$]{Clemens}.
Matters become slightly worse if we adopt $(\Theta_0+V_{\sun})/R_0 = 30.57\;\galunits$ estimated from star forming regions by \citet{Reid2014},
as this implies $V_{\sun} \approx 21\;\kms$.

In contrast, the fourth quadrant fit yields a higher LSR velocity of $\Theta_0 \approx 232\;\kms$.  
For the adopted $R_0$, this is consistent with the proper motion of Sgr A* if the solar motion is $V_{\sun} \approx 10\;\kms$.
Similarly, the star forming regions imply $V_{\sun} \approx 13\;\kms$.
These are plausible values consistent with many independent determinations \citep{Binney2010,MB2010,Schonrich2010,Sharma2014}.

At present, the true value of $\Theta_0$ seems systematically uncertain at the $\sim 10\;\kms$ level.
Indeed, this particular issue is greatly complicated by local gradients in the surface density and rotation curve in the immediate
vicinity of the sun \citep{OM98}.  The former should vary as we encounter the Perseus arm slightly outside the solar radius \citep{ravespirals}.
This can, in turn, cause a sudden change in $V(R)$.  
Consequently, it does not seem possible to improve much on the present models without a great deal more information
than currently available.  

\section{Conclusions}
\label{sec:concs}

We have applied the MDAR calibrated by external galaxies \citep{MDacc,GalRev} to the terminal velocity curves observed
in the first \citep{Clemens} and fourth \citep{luna,MGD} quadrants to construct mass models for the Milky Way.
These models provide a non-parametric numerical estimate of the surface density profile of the stellar disk $\Sigma_d(R)$.   
This provides a first step beyond the simple assumption of an exponential disk.

The Galactic disk inferred here is maximal and favors a short scale length.
The precise value of the scale length depends on the portion of the disk that is fit, potentially relieving the tension between apparently
discrepant determinations.  The scale length itself, while useful, is less fundamental than the pattern of structure to which it is fit.

The bumps and wiggles observed in the terminal velocities corresponds well to known spiral arms.
The arms are massive and have the expected effect on the observed kinematics.
These conclusions are independent of the bulge fraction or assumptions about the disk thickness,
though these do have a small effect the details of individual models.

The Milky Way appears to be a normal spiral galaxy.
It obeys scaling relations like the Tully-Fisher relation, the size-mass relation, and the disk maximality-surface brightness relation.
It is somewhat compact for its stellar mass, but resides well within the large intrinsic scatter of the size-mass relation.

In comparing the Milky Way to other galaxies, it is important to compare equivalent measures.
It is sometimes claimed that the Milky Way deviates from the Tully-Fisher relation,
but these claims seem to stem largely from using $\Theta_0$ as a proxy for other quantities.
The LSR velocity $\Theta_0$ is neither the peak of the rotation curve ($V_p > \Theta_0$) nor the outer, quasi-flat velocity ($V_f < \Theta_0$).
Though the difference between these quantities seems small, the scatter in the Tully-Fisher relation is very tight \citep{MS15}
so any difference is readily apparent.  

The distribution of stellar mass in the Galactic disk inferred here from the radial force is consistent with that inferred independently
from the vertical force \citep{BR13}.  Indeed, the vertical force is correctly predicted from our models with no adjustment.
The only point of disagreement between the two is where they probe different parts of the Galaxy, emphasizing the importance
of local structures like spiral arms.

One consequence of the bumps and wiggles in the terminal velocity curves and their correspondence to variations in the stellar surface density
is that the gradient of the rotation curve $dV/dR$ fluctuates on the scale of hundreds of parsecs.  This fluctuation is not subtle, having rms
amplitude $\langle$$|$$dV/dR$$|^2$$\rangle$$^{1/2} \approx 14\;\galunits$ in the fourth quadrant. 
This has important implications for Jeans analyses, which frequently invoke the flatness of the rotation curve to ignore terms involving its gradient.
Such approximations are unlikely to be adequate, especially to the analysis of the upcoming Gaia data.  
Indeed, it is not even adequate to assume a finite slope:  $dV/dR$ switches signs repeatedly on kpc scales.

The distribution of dark matter can be inferred from our models.
The amplitude of the mass discrepancy locally is $D(R_0) \approx 1.5$, leading to a spherically averaged
density of dark matter in the solar neighborhood $\rho_{0,DM} \approx 0.009\;\voldens$ ($0.34\;\gevcc$).
These values are likely accurate to 20\%, though we caution that systematic errors outweigh random ones.

More generally, the detailed radial distribution of dark matter can be empirically inferred from the models.
This is not precisely equivalent to any of the common halo forms, but can be tolerably approximated by an adiabatically compressed NFW halo.
The compression is important to reconciling the Milky Way halo with the mass-concentration relation expected in \LCDM,
and may help alleviate (though not solve) the too big to fail problem.
We do not, however, attempt to fit the inner 3 kpc, where the cusp-core problem persists.

As the only theory that predicted the MDAR, the the successes of these models can be interpreted as successes of MOND.
However, we identify a modest but apparently real tension between the radial and vertical forces in MOND: when the radial force is well fit,
the vertical force is over-predicted. This may be a genuine problem for MOND, but it may also be a hint about the deeper
theory underlying the observed phenomena.  We should keep an open mind about the underlying cause of the MDAR,
which may be a hint about the nature of dark matter as well as modified dynamical laws.

Irrespective of the underlying cause of the MDAR, the models presented here are based on an empirical calibration thereofof.
As such, they are empirically valid.  While the MDAR phenomenon deserves a better explanation than is currently available,
there is nothing to preclude us from using this information in our exploration of the Galaxy.

\acknowledgements  I thank Benoit Famaey, James Binney, Heather Morrison, Joss Bland-Hawthorn, Ivan Minchev, Giacomo Monari,
and Paul Harding for conversations on Galactic structure, and Jerry Sellwood and Harley Katz for work on adiabatic compression. 
This publication was made possible through the support of a grant from the John Templeton Foundation.
The opinions expressed in this publication are those of the author and do not necessarily reflect the views of the John Templeton Foundation.

\bibliography{ms_arxiv}
\bibliographystyle{apj}

\end{document}